\documentclass[fleqn,usenatbib]{mnras}

\usepackage{newtxtext,newtxmath}

\usepackage[T1]{fontenc}

\DeclareRobustCommand{\VAN}[3]{#2}
\let\VANthebibliography\thebibliography
\def\thebibliography{\DeclareRobustCommand{\VAN}[3]{##3}\VANthebibliography}

\usepackage{graphicx}	
\usepackage{amsmath}	

\usepackage{amssymb}	

\newcommand{\tb}{} 
\newcommand{\tr}{} 

\title[LAE to 21\,cm-line using GANs]{Predicting 21\,cm-line map from Lyman $\alpha$ emitter distribution with Generative Adversarial Networks}

\author[S. Yoshiura et al.]{
Shintaro Yoshiura,$^{1,2}$\thanks{E-mail: syoshiura@unimelb.edu.au}
Hayato Shimabukuro,$^{3}$
Kenji Hasegawa$^{4}$
and Keitaro Takahashi$^{5,6}$
\\
$^{1}${Mizusawa VLBI Observatory, National Astronomical Observatory Japan, 2-21-1 Osawa, Mitaka, Tokyo 181-8588, Japan}\\
$^{2}$The University of Melbourne, School of Physics, Parkville, VIC 3010, Australia\\
$^{3}$South-Western Institute for Astronomy Research (SWIFAR), Yunnan University (YNU), Kunming 650500, People's
Republic of China\\
$^{4}$Department of Physics and Astrophysics, Nagoya University Furo-cho, Chikusa-ku, Nagoya, Aichi 464-8602, Japan\\
$^{5}$Faculty of Science, Kumamoto University, 2-39-1 Kurokami, Kumamoto 860-8555, Japan\\
$^{6}$International Research Organization for Advanced Science and Technology, Kumamoto University, 2-39-1 Kurokami, Kumamoto 860-8555, Japan
}

\date{Accepted XXX. Received YYY; in original form ZZZ}

\pubyear{2020}

\begin{document}
\label{firstpage}
\pagerange{\pageref{firstpage}--\pageref{lastpage}}
\maketitle

\begin{abstract}
The radio observation of 21\,cm-line signal from the Epoch of Reionization (EoR) enables us to explore the evolution of galaxies and intergalactic medium in the early universe. However, the detection and imaging of the 21\,cm-line signal are tough due to the foreground and instrumental systematics. In order to overcome these obstacles, as a new approach, we propose to take a cross correlation between observed 21\,cm-line data and 21\,cm-line images generated from the distribution of the Lyman-$\alpha$ emitters (LAEs) through machine learning. In order to create 21\,cm-line maps from LAE distribution, we apply conditional Generative Adversarial Network (cGAN) trained with the results of our numerical simulations. We find that the 21\,cm-line brightness temperature maps and the neutral fraction maps \tr{can be reproduced with correlation function of 0.5 at large scales $k<0.1~{\rm Mpc}^{-1}$}. \tb{Furthermore, we study the detectability of the the cross correlation assuming the the LAE deep survey of the Subaru Hyper Suprime Cam, the 21\,cm observation of the MWA Phase II and the presence of the foreground residuals. We show that the signal is detectable at $k < 0.1~{\rm Mpc}^{-1}$ with 1000 hours of MWA observation even if the foreground residuals are 5 times larger than the 21\,cm-line power spectrum.} Our new approach of cross correlation with image construction using the cGAN can not only boost the detectability of EoR 21\,cm-line signal but also allow us to estimate the 21\,cm-line auto-power spectrum.
\end{abstract}




\section{Introduction}\label{Sec:intro} 

One of the milestones in the history of the Universe is the epoch of reionization (EoR). The ultraviolet (UV) photons emitted from ionizing sources such as early galaxies ionize the hydrogen atoms in the intergalactic medium (IGM) and generate ionized bubbles around ionizing sources. The morphology and topological property of ionized regions depends on the nature of the ionizing sources. However, the dominant source of ionizing photons still remains unknown. The redshifted cosmological 21\,cm-line from the neutral hydrogen atom is one of promising tools to probe the morphology of ionized bubbles in the IGM during the EoR.

Currently, low-frequency radio telescopes such as the the Giant Metrewave Radio Telescope EoR Experiment \citep[GMRT,][]{2013MNRAS.433..639P} the Donald C. Backer Precision Array for Probing the Epoch of Reionization \citep[PAPER,][]{2010AJ....139.1468P}, the Murchison Widefield Array \citep[MWA,][]{2013PASA...30....7T} and the LOw Frequency ARray \citep[LOFAR,][]{2013A&A...556A...2V} are operating to detect the 21\,cm-line signal from the EoR. These ongoing radio telescopes have provided upper limits on the 21\,cm-line power spectrum, and these upper limits have been gradually updated thanks to sophisticated analysis \citep[e.g][]{2019ApJ...883..133K,Trott2020DeepObservations2,2020MNRAS.493.1662M}

However, we are facing many difficulties in detecting the 21\,cm-line. Specifically, observations of the 21\,cm-line from the EoR are obstructed by the bright foreground contamination and observational systematic errors. The foreground contamination is mainly due to the synchrotron emission from our Galaxy and extragalactic radio sources and is about 4 orders of magnitude larger than the 21\,cm-line signal. {Furthermore}, the observational data are polluted by various systematics such as Earth's ionosphere, the beam-shape error and the radio frequency interference. Thus, it is required to perform a foreground removal/avoidance with a high accuracy and overcome these systematics adequately in order to detect the 21\,cm-line signal.

The cross correlation between the 21\,cm-line and other emission lines is one of {the most} powerful methods to reduce the effects of contamination from the foreground and systematics because the foreground and systematics should not correlate with the partner lines of the cross correlation. High-$z$ galaxies have been studied as the partner of the cross correlation in previous works  \citep[e.g.][]{2009ApJ...690..252L,2013MNRAS.432.2615W,2014MNRAS.438.2474P,2016MNRAS.459.2741S,2017ApJ...836..176H,2019MNRAS.tmp.2236M,2017ApJ...846...21F,2017ApJ...848...52H,2017ApJ...846...21F,2019BAAS...51c..57H,2020MNRAS.492.4952V,2020MNRAS.494..703W,Kubota2020MNRAS.494.3131K,2020MNRAS.496..581H}. Since the ionizing photon emitted from high-$z$ massive galaxies generate large ionized regions around them, the 21\,cm-line signal negatively correlates with galaxies at large scales. In addition to the negative correlation at large scales, the 21\,cm-line signal can positively correlate with galaxies at small scales due to matter density fluctuation. 

In \citet{2018MNRAS.479.2754K}, we have studied the Lyman-$\alpha$ emitters (LAEs) as a partner of the cross correlation. The LAE is a kind of high-$z$ galaxies which emit the strong Lyman-$\alpha$ line. So far, Subaru Hyper Surpreme Cam (HSC) has detected a large number of LAEs in large survey fields  at $z>5.7$ \citep{2018PASJ...70S..13O,2018PASJ...70S..14S,2018PASJ...70S..16K,2021arXiv210402177O} and further surveys at higher redshifts are ongoing. We have found the cross power spectrum between the 21\,cm-line and Lyman-$\alpha$ emitters can be detected by combining the MWA, LAE surveys by the Subaru HSC and a follow-up by the Prime Focus Spectrograph \citep[PFS,][]{2014PASJ...66R...1T,2016SPIE.9908E..1MT} to determine precise redshifts of the LAEs.

\tb{Regarding the analysis of 21\,cm-line at the EoR, the applications of machine learning technique, in particular artificial neural network (ANN), have been suggested  \citep[e.g.][]{2017MNRAS.468.3869S,2019arXiv190404106D,2018arXiv180502699G,2019ApJ...880..110L,2017ApJ...848...23K,2018arXiv181109141J,2018MNRAS.475.1213S,2019MNRAS.483.2524H,2019arXiv190707787H,2019MNRAS.485.2628L,2020arXiv200208238S,2020MNRAS.493.5913L}. Recently, the convolutional neural network (CNN) which is one of the deep learning algorithms is recognized as a powerful tool for foreground removal. For example, in \cite{2020arXiv201015843M}, they used the CNN with a U-net architecture to retrieve the 21cm map from the foreground contaminated 21cm map. Furthermore, in the context of cosmology, the deep learning algorithms are actively used to solve the problem of image translation. For instance, the U-net architecture is used to predict the galaxy distribution from the matter density distribution \citep{2019arXiv190205965Z} and to predict  non-linear large scale structure from matter density distribution generated by using Zel’dovich approximation \citep{2019PNAS..11613825H}. In addition, the conditional Generative Adversarial Networks (cGAN) method has also been used in making independent images of weak lensing convergence maps  \citep{2019ComAC...6....1M}, the cosmic neutral hydrogen distribution \citep{2019arXiv190412846Z} and 21\,cm-line \citep{2020MNRAS.493.5913L}. The cGAN method is not only used for creating new images but also for image-to-image translation such as adding the effect of dark matter annihilation feedback (DMAF) to gas density map \citep{2019MNRAS.490.3134L}, deriving a map of the gas pressure distribution from dark matter density \citep{2019MNRAS.487L..24T}, de-noising an observed weak lensing mass map \citep{2019PhRvD.100d3527S} and separating different emission lines from line intensity mapping images \citep{2020MNRAS.496L..54M}.}

\tb{In this work, we propose a new cross correlation approach to detect the 21\,cm-line signal using the cGAN.} First, we predict 21\,cm-line maps from the observed spatial distribution of LAEs with the cGAN, which is one of the ANN techniques. Specifically, to predict 21\,cm-line maps from LAE distribution, we use an image-to-image translator developed in \cite{Phillip2016} based on the cGAN technique. As mentioned above, the 21\,cm-line signal correlates with the LAE distribution at scales of a wide range, and therefore it would be possible to predict the 21\,cm-line maps from the observed LAE distribution if the network has learned the 21cm-LAE cross correlation properly in advance. The predicted 21\,cm-line maps can be used as the partner of the cross correlation to not only boost the detectability of 21\,cm-line signal but also estimate its auto-correlation which cannot be obtained by the conventional cross-correlation method with galaxies.

We show the feasibility of our new method using simulations. First, we develop a network with training datasets, which consist of input data (LAE spatial distribution) and {target} data (21\,cm-line map). We employ the result of two types of numerical simulations of galaxy formation and reionization to prepare the training dataset. Then we evaluate the performance of the network with test dataset which are also obtained from the numerical simulations. Furthermore, we investigate the detectability of the cross correlation between the predicted 21\,cm-line maps and the {mock} observed 21\,cm-line data assuming 21\,cm-line observations by the MWA and Subaru HSC. \tb{Moreover, we discuss the required levels of foreground removal for the detection.} Finally, we argue the estimation of the 21cm auto-power spectrum using the cross correlation.

This paper is structured as follows. In Sec.~\ref{Sec:21cm}, we describe the 21\,cm-line and our reionization simulation model. In Sec.~\ref{Sec:NN}, we introduce the cGAN method and a training dataset used in this work.  
In Sec.~\ref{Sec:ANA}, we outline our method for evaluation of the detectability of the 21\,cm-line. In Secs.~\ref{Sec:res} and \ref{Sec:dis}, we show our results and give a discussion. Finally, we summarise our work in Sec.~\ref{Sec:sum}.

\section{21cm-line and Simulation}\label{Sec:21cm} 

In this work, we attempt to construct 21\,cm-line maps from spatial distribution of LAEs. The redshifted 21\,cm-line, emitted from neutral hydrogen, is measured as the brightness temperature which is described as \citep[e.g.][]{Furlanetto:2006jb},
\begin{eqnarray}
\delta T_b &=& 27 x_{\rm HI} (1+\delta_{\rm m})\left( 1-\frac{T_{\rm CMB}}{T_S}\right) \nonumber \\
&&\times  \left( \frac{\Omega_b h^2}{0.023}\right) \left( \frac{0.15}{\Omega_{m}h^2}\frac{1+z}{10} \right)^{1/2} \rm mK,
\label{eq:21cm} 
\end{eqnarray}
where $x_{\rm HI}$ is the neutral fraction, $\delta_{\rm m}$ is the matter density fluctuation, $T_{\rm CMB}$ is the CMB temperature, $T_S$ is the spin temperature, $h$ is the {Hubble constant}, $\Omega_b$ and $\Omega_m$ are the {present day} density parameter of baryon and matter, respectively. We assume the IGM is sufficiently heated and $T_{\rm S} \gg T_{\rm CMB}$. This assumption is reasonable since we focus on lower redshift $z=6.6$. 

For modeling the 21\,cm-line and the LAE distribution, we employ a radiative transfer (RT) numerical simulation and a semi-numerical simulation for solving the ionization. {By comparing the networks developed from two types of simulation, we discuss the effect of the model of ionization to the accuracy of signal prediction.} The LAE distribution is evaluated by solving a Lyman-$\alpha$ 1D RT from the data. We describe the models below.

\subsection{Radiative Transfer}\label{sSec1:}

We briefly describe our RT reionization simulation which is identical to that studied in \cite{2018MNRAS.479.2754K}.

Our simulation employs the matter density fields obtained from a massive cosmological N-body simulation \citep{2012arXiv1211.4406I, 2009PASJ...61.1319I}. The RT equations are solved in a box of $\rm 160~Mpc^3$ gridded in $256^3$ cells. The area of the box, $160^2~\rm deg^2$, approximately corresponds to $1~\rm deg^2$ at $z=6.6$. As the model of ionizing sources, we employ the result of a radiative hydrodynamic (RHD) simulation \citep{2013MNRAS.428..154H,2016arXiv160301961H}. The RHD simulation regulates the star formation by taking into account the UV feedback and the supernovae feedback. The clumping factor depends on the matter density and the neutral fraction. It is worth noting that the IGM ionization history of the model is consistent with the constraints by quasar spectra \citep{2006AJ....132..117F} and Thomson scattering optical depth to the CMB \citep{2016A&A...596A.108P}.

We assume the HSC deep survey at $z$=6.6. However, in order to prepare a large number of training datasets, we use the IGM data {at 20 different redshifts in a range of $6.3<z<7.4$.} For the calculation of the LAE distribution, {we solve the Lyman-$\alpha$ radiative transfer using the IGM data} and use the halo distribution at $z=6.6$ to avoid the evolution of the halo mass function.

\subsection{Semi-Numerical Model}\label{sSec2:}

We use a simple model calculating the reionization process with a semi-numerical scheme to compare the performance of the networks. {The halo distribution and the matter density field are identical to those used in the RT model. The semi-numerical method used in this work is similar to that of 21cmFAST \citep{2011MNRAS.414..727Z, 2011MNRAS.411..955M}, while the latest version of 21cmFAST \citep{2019MNRAS.484..933P} adopts a more realistic model.}

First, the number of ionizing photons emitted from a halo into IGM is modeled as
\begin{eqnarray}
\dot{N}_{\rm ion} = f_{\rm esc,c} \left( \frac{M_{\rm halo}}{10^{10}}\right)^{\alpha_{\rm esc}} \dot{N}_{\rm ion,int},
\end{eqnarray}
where $f_{\rm esc,c}$ is the escape fraction of ionizing photons, $M_{\rm halo}$ is the halo mass, $\alpha_{\rm esc}$ controls a mass dependence of the $f_{\rm esc,c}$. The number of ionizing photons produced in a halo is given as
\begin{eqnarray}
\dot{N}_{\rm ion,int} \propto (10^{-0.58M_{\rm UV}}),
\end{eqnarray}
where we convert $M_{\rm halo}$ to the UV magnitude $M_{\rm UV}$ based on \cite{2014MNRAS.440..731S}.

The ionization state is determined by comparing the number of ionizing photons and the number of neutral hydrogen atoms within a sphere of radius $R$ centred at a cell. We start the comparison from $R=R_{\rm max}$ and eventually decrease the $R$ to the size of a cell. Here $R_{\rm max}$ is {the maximum mean free path of ionizing photon} \citep{2011MNRAS.414..727Z}. If the number of ionizing photons is larger than the number of neutral hydrogen atoms, the central cell of the sphere is regarded as ionized. In fact, we assume the neutral fraction is $10^{-4}$, rather than exactly zero, inside the ionized bubble. We normalize $\dot{N}_{\rm ion,int}$ so that the neutral fraction is 0.2 for the model with $f_{\rm esc,c} = 0.001$, $\alpha_{\rm esc} = -1.0$ and $R_{\rm max} = 30\rm Mpc$ at $z$=6.6.

For preparing the training and test datasets, we vary the parameters in ranges of $f_{\rm esc,c} = [0.001,0.5]$, $\alpha_{\rm esc} = [-1.0,0.0]$ and $R_{\rm max} = [5,50]$. We label the model with $-1.0<\alpha_{\rm esc}<-0.5$ as $S_l$ and the model with $-0.5<\alpha_{\rm esc}<0$ as $S_h$.

\subsection{LAE model}

The distribution of observable LAEs is obtained by solving 1D RT of Lyman-$\alpha$ photons using the IGM data described above. The model of LAEs is identical to the model used in \cite{2018MNRAS.479.2754K,2018MNRAS.479.2767Y}. The intrinsic luminosity of Lyman-$\alpha$ is based on the model of the RHD simulation and given as $L_{\alpha} = 10^{42} (M_{\rm halo}/10^{10})^{1.1}$[erg/s]. We use the line profile model of \cite{2018MNRAS.477.5406Y}, and derive the transmission rate by solving the optical depth through the IGM along the line of sight (LoS). We identify halo as LAEs if the luminosity is larger than $4.2 \times 10^{42}~{\rm [erg/s]}$ and $2.5 \times 10^{42}~{\rm [erg/s]}$ for the HSC deep and ultra-deep surveys, respectively.

The LAE model has free parameters such as the Lyman-$\alpha$ escape fraction $f_{\alpha}$, the galactic wind velocity $V_{\rm out}$, and the HI column density within the galaxy $N_{\rm HI}$. In practice, the LAE parameters can be calibrated so that the LAE distribution is consistent with latest observations. Although the calibration of the LAE parameters depends on the model of reionization, we fix the LAE parameters which are consistent with that of \cite{2018MNRAS.479.2754K} for simplicity: $f_{\alpha} = 0.3$, $V_{\rm out} = 150~{\rm [km/s]}$ and $N_{\rm HI} = 10^{19}~{\rm [cm^{-2}]}$. We leave the calibration of the LAE model considering the reionization model for future works. It should be worth to note that \cite{2018PASJ...70...55I} has proposed stochastic models of LAE, which can explain various statistical properties of LAEs measured by the HSC survey.

\section{Image to Image Translator}\label{Sec:NN} 

\tr{In order to construct maps of the 21\,cm-line and the neutral fraction from the LAE distribution, we use the method proposed in \cite{Phillip2016}. This method is based on the conditional GAN. In the original literature, they have shown great success for the image-to-image translation.} In this section, we describe the methodology and training dataset.

\subsection{Generative Adversarial Networks}

The GAN, originally developed in \cite{2014arXiv1406.2661G}, consists of two networks, a generator and a discriminator. The generator aims to produce fake images which are not distinguishable from real images. The discriminator attempts to distinguish real images and fake images created by the generator.

In \cite{Phillip2016}, the authors have applied the conditional GAN (cGAN) to the image-to-image translation problem, and the code has been made public\footnote{https://phillipi.github.io/pix2pix/}\footnote{In this work, we use a port of the code\\ https://github.com/affinelayer/pix2pix-tensorflow.}.
\tb{This is a supervised learning method where pairs of the target and input images are explicitly learned\footnote{\tb{A more simple method such as CNN with a U-net architecture can be used for the same purpose. However, comparing the efficiency of deep learning methods is out of scope of this work. We leave it for future works.}}. In this work, we employ this architecture.} Below, we describe the methodology briefly \tb{following \cite{Phillip2016}. See the original literature }for more details and examples. 

The generator $G$ and discriminator $D$ \tb{attempt to minimize and maximize} the objective $\mathcal{L}_{\rm pix}$ respectively,
\begin{eqnarray}
\tb{G^* = {\rm arg\,min_G\,max_D} \mathcal{L}_{\rm pix}.}
\label{eq:pix2pix0} 
\end{eqnarray}
The objective function is given as
\begin{eqnarray}
\mathcal{L}_{\rm pix} = \mathcal{L}_{\rm cGAN}(G,D)+\lambda \mathcal{L}_{\rm L1}(G)
\label{eq:pix2pix} 
\end{eqnarray}
where a hyperparameter $\lambda$ works as the weight of the term. The first term is a general objective for the cGAN, and it is given by
\begin{eqnarray}
\mathcal{L}_{\rm cGAN}(G,D) &=& {\mathbb{E}}_{x,y}[\log D(x,y)]\nonumber \\
&+&{\mathbb{E}}_{x}[\log(1-D(x,G(x)))],
\label{eq:cGAN} 
\end{eqnarray}
where $\mathbb{E}[a]$ represents the expectation value of $a$, $x$ is an input image and $y$ is a real image which is the genuine pair image of the $x$. The generator, $G$, makes a fake image, $G(x)$, from an input $x$. The discriminator, $D(x,z)$, judges if a given image $z$ is $y$ or a fake image $G(x)$ and returns a value from 0 (fake) to 1 (real).

In \cite{Phillip2016}, they have introduced $L_1$ as the second term in Eq.~(\ref{eq:pix2pix}). The $L_1$ is given as
\begin{eqnarray}
\mathcal{L}_{\rm L1}(G)={\mathbb{E}}_{x,y}(||y-G(x)||_1).
\label{eq:L1} 
\end{eqnarray}
The term represents the difference of the generated fake image $G(x)$ and the real image $y$. By introducing the $L_1$ term, the generator is trained to generate images close to real images and to deceive the discriminator. In \cite{Phillip2016}, they have argued that the $L_1$ can reduce the artificial behavior of output images.

In our context, the input image $x$ is an image of LAE distribution and $y$ is the real image of the 21\,cm-line signal corresponding to the $x$. Both are calculated from our numerical simulation in this paper. The $G(x)$ is a fake image of 21\,cm-line signal generated from the LAE distribution $x$. The discriminator $D$ is given a pair of images of the LAE distribution and 21\,cm-line map and judges if the latter is a real or fake image.

\tb{Next, we describe the architecture briefly. See \citep{Phillip2016} and a document of this implementation\footnote{https://affinelayer.com/pix2pix/} for details. The \tb{generator architecture is based on a U-net architecture} \citep{2015arXiv150504597R}. The architecture used in this work is identical to the one used in the original paper except the number of convolution layers of generator $N_{\rm d}$ and the number of filters of generator $f_{\rm g}$ and discriminator $f_{\rm d}$ as we treat these values as the hyperparameters. In the generator architecture, input LAE image is processed through $N_{\rm d}$ down sampling convolution layers and $N_{\rm d}$ up sampling de-convolution layers with skip connections between $i$-th convolution layer and $N_{\rm d}-i$-th de-convolution layer. The size of the filter is $4\times4$ and the stride is 2. The discriminator architecture consists of 5 down sampling convolution layers. The pair of images is down sampled through 4 layers and output of the fifth layer is a $30 \times 30$ image which is used to judge the image to be real or fake.}

\subsection{Datasets and Training}\label{sec:3.2}

Here we describe the datasets used to train and test the network. They are obtained from the numerical simulations described in the previous section.

First, because the pixel value of input and output images needs to be in a range of [0-255], we normalize the 21cm brightness temperature images by $20~{\rm [mK]}$ and the LAE images by $L_{\rm max} = 2.1 \times 10^{43}~\rm [erg/s]$. In order to improve the efficiency of training, we classify the images into three groups based on the average neutral fraction such as $x_{\rm HI}<0.1$, $0.1<x_{\rm HI}<0.3$, and $0.3<x_{\rm HI}<0.5$.

We label the dataset as follows. We refer to the RT model and semi-numerical model as $H$ and $S$, respectively. For $S$ models, we label the model with $-1.0 < \alpha_{\rm esc} < -0.5$ and $-0.5 < \alpha_{\rm esc} < 0$ as $S_l$ and $S_h$, respectively. The 3 bins of the average neutral fraction of $x_{\rm HI} < 0.1$, $0.1 < x_{\rm HI} < 0.3$, and $0.3 < x_{\rm HI} < 0.5$ are referred as 1, 2 and 3, respectively. Thus, for example, a network trained using the RT simulation data of the neutral fraction bin of $0.1 < x_{\rm HI} < 0.3$ is referred as ``$H_{2}$''. Note that the HSC deep survey is assumed in most cases and we add $\rm UD$ when the HSC ultra-deep survey is assumed such as ``$S_{l,2} \rm UD$''.

\tb{We build 7 reionization models ($H_1$, $H_2$, $H_3$, $S_{l,2}$, $S_{l,3}$, $S_{h,2}$, $S_{h,3}$) and 2 LAE survey models (deep and ultra deep). Thus, we use 14 datasets of 21\,cm-line and LAE images for the training. As the original work, the networks are trained using Adam solver \citep{2014arXiv1412.6980K}.} \tr{When we create the 21\,cm-line images using the trained network, the input LAE model is matched with the network except in section 5.4 where the input LAE distribution is based on the $H_{2}$ model. }

We need a large number of training images to train the network effectively. However, we have only one realization of RT simulation due to its high computational cost. Thus, we increase the number of quasi-independent images from 1 simulation box as follows. Because the redshift uncertainty of the Lyman-$\alpha$ survey of the Subaru HSC is $\Delta z = 0.1$ which roughly corresponds to $40~{\rm Mpc}$, we divide the simulation box into 4 slices of $(160~{\rm Mpc})^2 \times 40~{\rm Mpc}$. By doing this along the 3 axes of the box, we obtain 12 slices which correspond to the sky area of $1~{\rm deg}^2$ and the redshift width of $0.1$. Further, we randomly shift the center of the maps and create quasi-independent images to obtain 500 and 100 images as training and test dataset, respectively.

If the training \tb{dataset includes identical images to test images, the network loses the versatility.} To avoid this and keep the independence between training and test data, we use slices divided along x and y directions as training dataset and those divided along z direction as test dataset. We list the number of original training and test images in table.~\ref{table:model}.

\tb{In order to tune the hyperparameters, we  perform a validation. One validation dataset consists of 50 images selected from the training dataset of $S_{l,2}$ and the rest of 450 images are used for training. \tb{We note that just the $S_{l,2}$ model is used for the cross validation due to our limited computational resources although the cross validation should be performed for each reionization model.} By choosing the validation image randomly, we make \tb{10} independent \tb{cross} validation datasets. The validation is performed by varying the hyperparameters such as $\lambda$, the number of convolution layers $N_{\rm D}$, the number of filters of first layer of the generator $f_{\rm g}$ and the discriminator $f_{\rm d}$, batch size $b_{\rm s}$, the initial learning rate for Adam $Lr$ and the momentum term of Adam $\beta$. We fix the other parameters (e.g. the number of layers of discriminator $N_{g}=5$, a dropout rate of 50\% and the second momentum term of Adam $\beta_2=0.999$) to the fiducial values following previous works. We train the networks for 500 epochs. \tb{To speed up our validation in limited computation resources, we proceed it through 3 steps. First, we perform the validation} using only one of the validation datasets. We compare the real-space cross-correlation coefficient (CCe) between target images and output images for 100 parameter sets, where the parameter values are randomly chosen within a prior distribution. From the top 20 parameter sets, we identify a likely range of parameter values. Then, making 50 parameter sets from the range, we perform the second validation. Finally, we perform a validation using all validation datasets for the 5 best parameter sets in the among the 150 parameter sets, and choose the best parameter set which achieves the maximum CCe. The initial prior distribution and best parameter sets are listed in Table.~\ref{tab:params}.}

\tb{For the $H_1$ model, however, \tb{the network trained with the best hyperparameters} generates completely zero 21cm signals from any input LAE maps. This indicates the different hyperparameters are preferred for the $H_1$ model. For an experiment, we train the network with the best parameters but with the $\lambda=100$. Then the resultant 21cm maps have non-zero values. Even with $\lambda=100$, the correlation coefficient between target and output images is low \tb{$r<0.2$}, and the parameter sets does not work correctly for reconstructing the neutral fraction images. As the cGAN to the $H_1$ model is unstable, we do not perform further hyperparameter search for the $H_1$ model. }

\begin{table}
  \begin{tabular}{ccccccc}
   \hline \hline
  model&$H_{1}$& $H_{2}$  &$H_{3}$ & & &   \\
  $N_{\rm train}$    & 188  &   100  &  76   &&&  \\
  $N_{\rm test}$    &  92   &   52  & 40   &&&\\
  \hline
  model & $S_{l,2}$ & $S_{l,3}$  &$S_{h,2}$ & $S_{h,3}$ &&  \\
  $N_{\rm train}$    &  140  &  203   &     267& 157    &     &   \\
  $N_{\rm test}$    &  68   & 116    & 133   & 72   &    &   \\
   \hline
  \end{tabular}
 \caption{\label{table:model}The number of the training datasets $N_{\rm train}$ and the test datasets $N_{\rm test}$. The RT simulation and the semi-numerical models are labeled as $H$ and $S$, respectively. The subscript $i$ ($i = 1,2$ and $3$) indicates the neutral fraction ($x_{\rm HI} < 0.1$, $0.1 < x_{\rm HI} < 0.3$, and $0.3 < x_{\rm HI} < 0.5$, respectively). In the semi-numerical model, $l$ and $h$ indicate the power-law index of the escape fraction is lower and higher than -0.5, respectively.}
\end{table}

\section{Statistical Analysis}\label{Sec:ANA}

Here, we describe the data analysis method used in this work. For the evaluation of an accuracy of our network, we calculate the correlation coefficient in real space, which is given as
\begin{eqnarray}
r = \frac{ \sum_i^N (a_i-\bar{a})(b_i-\bar{b}) }{\sqrt{\sum_i^N (a_i-\bar{a})^2}\sqrt{\sum_i^N (b_i-\bar{b})^2}},
\end{eqnarray}
where $N$ is the number of pixels, $a_i$ is $i$th pixel value of output images, $b_i$ is $i$th pixel value of real images and $\bar{a}$ represents the averaged value.

A major tool to study the 21\,cm-line is the auto power spectrum, which is given by 
\begin{eqnarray}
P (|{\bf{k}}|) = (2\pi)^2\delta_{\rm D} ({\bf{k}}-{\bf{k}}') \langle \delta({\bf{k}}) \delta({\bf{k}')}\rangle,
\end{eqnarray}
where ${\bf{k}}$ is wavenumber in 2D Fourier space, $\delta_D$ is Dirac's delta function, $\delta$ is the fluctuation in 2D Fourier space.

We also use the 21cm-LAE correlation coefficient which is given as,
\begin{eqnarray}
r_k^{21,\rm LAE}(k) = \frac{P_{21,\rm LAE}(k)}{\sqrt{P_{21}(k)P_{\rm LAE}(k)}},
\label{eq:cross_21-LAE}
\end{eqnarray}
where $P_{21}$ is the 21cm power spectrum and $P_{\rm LAE}$ is the power spectrum of LAE distribution. The 21cm-LAE cross power spectrum is given as,
\begin{eqnarray}
P_{21,\rm LAE} (|{\bf{k}}|) = (2\pi)^2\delta_{\rm D} ({\bf{k}}-{\bf{k}}') \langle \delta({\bf{k}})\delta_{LAE}({\bf{k}')}\rangle,
\end{eqnarray}
where $\delta_{\rm LAE}$ is the fluctuation of the LAE distribution.

In this work, we attempt to calculate the cross correlation between the target 21\,cm-line image and the output 21\,cm-line image, which is given as 
$P^{T,O}_{X} (|{\bf{k}}|) = (2\pi)^2\delta_{\rm D}({\bf{k}}-{\bf{k}}') \langle \delta_{\rm tar} ({\bf{k}})\delta_{\rm out} ({\bf{k}}) \rangle$,
where $\delta_{\rm tar}$ and $\delta_{\rm out}$ are the fluctuation in the target image and the output image, respectively.  
The correlation coefficient is given as 
\begin{eqnarray}
r^{T,O}_k(k) = \frac{P^{T,O}_X(k)}{\sqrt{P_{\rm tar}(k)}\sqrt{P_{\rm out}(k)}},
\label{eq:rTO}
\end{eqnarray}
where $P_{\rm tar}$ and $P_{\rm out}$ indicates the auto power spectrum of the target image and the output image, respectively.

Ultimately, in the future work, the output image is used for the cross correlation with the observed 21\,cm-line. In this work, we assume target images as the observed 21\,cm-line and estimate the detectability of the cross power spectrum. We evaluate the error which is written as (e.g. \cite{2009ApJ...690..252L}),
\begin{eqnarray}
&{2}&\sigma^{2}_X(|{\bf{k}}|)  = \frac{1}{\epsilon \Delta k k S_{\rm area}/2\pi} \times
\left( P^{2}_X(|{\bf{k}}|) \right. \\
&& \left. + \left(P_{\rm obs}(|{\bf{k}}|) + N_{\rm obs}(|{\bf{k}}|) + P_{\rm FG}(|{\bf{k}}|)\right) \left(P_{\rm out}(|{\bf{k}}|) + N_{\rm pix}(|{\bf{k}}|)\right)\right) \nonumber,
\label{eq:error}
\end{eqnarray}
where $P_X$ is the cross power spectrum between the observed 21\,cm-line image and the output image, $P_{\rm obs}$ is the power spectrum of observed 21\,cm-line image \tb{and $P_{\rm FG}$ is the power spectrum of the foregrounds}. The number of samples in $k$ bin is ${\epsilon \Delta k k S_{\rm area}/2\pi}$ where $\Delta k$ is the width of k bin and $S_{\rm area}$ is the area of image. In practice, we replace the term by the number of ${\bf{k}}$ samples used in power spectrum calculation. $N_{\rm obs}$ is thermal noise, and $N_{\rm pix}$ is the error due to the network. 

\tb{
For the 21\,cm-line observation, we assume an observation by the MWA Phase II compact configuration \citep{2018PASA...35...33W}, and the thermal noise can be estimated by (e.g. \cite{2006ApJ...653..815M})
\begin{eqnarray}
N_{\rm obs}({\bf{k}}) = \left(\frac{\lambda_o^2}{A_{\rm e}}\right)^2 \left(\frac{T_{\rm sys}}{\sqrt{Bt}}\right)^2\frac{D^2_{\rm M}(z)}{n({\bf{k}})},
\end{eqnarray}
where $n({\bf{k}})$ is the number density of baselines in a $k$ bin, $\lambda_o$ is the observed wavelength, $A_{\rm e}=21.5 (\nu/150\rm ~MHz)^{-2}$[$\rm m^2$] is the effective area of antenna, $T_{\rm sys}=280 ((1+z)/7.5)^{2.3}$ [K] is the system temperature, $B=8.0~$MHz is the bandwidth, $t$ is the observation time and $D_{\rm M}(z)$ is the comoving distance to $z$. For estimation of baseline distribution $n({\bf{k}})$, we use the array distribution of MWA\footnote{http://www.mwatelescope.org/telescope/configurations/phase-ii}. }

\tb{
\tb{Foregrounds dominate the error described in Eq.~\ref{eq:error} without a high quality foreground removal. For quantitative discussion about the level of foreground removal  required to detect the 21cm signal, as} the reference foreground power spectrum, we use the foreground model including Galactic and extragalactic synchrotron emission motivated by \cite{2005ApJ...625..575S,2008MNRAS.389.1319J} which is given as 
\begin{eqnarray}
P_{\rm FG}(k) &=& T^2_{\rm GF} (k/k_{\rm ref})^{\alpha_{\rm GF}} (\nu/\nu_{\rm ref})^{\beta_{\rm GF}} \nonumber \\
&+& T^2_{\rm EGF} (k/k_{\rm ref})^{\alpha_{\rm EGF}} (\nu/\nu_{\rm ref})^{\beta_{\rm EGF}} \rm [K^2 Mpc^2]
\label{eq:fg}
\end{eqnarray}
where we set the amplitude of Galactic and extragalactic foreground emission  $T_{\rm GF}=T_{\rm EGF}=350$, the power law index of $\alpha_{\rm GF}=-2.7$ and $\alpha_{\rm EGF}=-1.1$, the spectral index of $\beta_{\rm GF}=-2.55$ and $\beta_{\rm EGF}=-2.00$, the reference scale of $k_{\rm ref}=0.05 ~\rm h\,Mpc^{-1}$ and the reference frequency of $\nu_{\rm ref}=180 ~\rm MHz$. As the extragalactic sources can dominate the diffuse emission at $k>0.1 \rm Mpc^{-1}$, we choose the reference scale of $k_{\rm ref}=0.05 ~\rm h\,Mpc^{-1}$ so that the Galactic foreground has the same amplitude as extragalactic foreground at the scale. The contamination from the extragalactic and Galactic free-free emission are weaker than the synchrotron emission, and therefore we ignore them. This foreground power spectrum\footnote{\tb{Our foreground power spectrum is roughly 40 [$\rm K^2$] at the reference scale and the amplitude is more or less consistent with previous works (\cite{2005ApJ...625..575S,2008MNRAS.389.1319J}). The amplitude of the foregrounds power spectrum is also motivated by the amplitude at lowest $k_{\parallel}$ shown in the cylindrical power spectrum of MWA observation such that $P_{\rm k}\approx 10^{12}$ [$\rm mK^2 h^{-3} Mpc^3$]  at \tb{$k_{\perp}=0.05~h\rm Mpc^{-1}$ and $k_{\parallel}=0.01~h\rm Mpc^{-1}$ after removal of bright extragalactic sources} \citep[e.g.][]{Barry2019ImprovingObservations2}. }} is more than 7 orders of magnitude larger than the 21\,cm-line power spectrum of $S_{l,2}$ model at the reference scale.}

The error on the cross power spectrum can be reduced by adding independent sample images. Thus, the error is evaluated as $\sigma_X/\sqrt{N}$, where $N$ is the number of samples. For example, \tb{there are four HSC deep fields and the total field of view is 27$\rm deg^2$. \tb{As our network reconstructs the 21cm line images of 1~$\rm deg^2$}, we set $N=27$ and $S_{\rm area}=1~\rm deg^2$. }

\section{Results}\label{Sec:res} 

We here show results of the 21\,cm-line image prediction by our cGAN network, the power spectrum and the 21cm-LAE cross power spectrum. We have 7 training datasets (i.e. $H_1$, $H_2$, $H_3$, $S_{l,2}$, $S_{l,3}$, $S_{h,2}$ and $S_{h,3}$) for deep and ultra deep surveys. Thus, 14 different networks are developed.

The structure of this section is as follows. We first check the property of data used for training in section \ref{sec:5.1}. Next, we compare the predicted images, their auto-power spectra and cross correlation coefficients with those of target images in sections \ref{sec:5.2} and \ref{sec:5.3}. In section \ref{sec:5.4}, in order to see the robustness of our method, we attempt to predict 21\,cm-line images of $H_2$ model using networks trained with datasets of other models. Finally, the detectability of cross-power spectrum between the observed 21\,cm-line image and predicted image is discussed in section \ref{sec:5.5}, and we attempt to estimate 21\,cm-line auto-power spectrum using the cross-power spectrum in \ref{sec:5.6}. Note that the correct model is identical to the model used for the training network in this section except in \ref{sec:5.4}.

\subsection{Training Data Property}\label{sec:5.1}

Before showing results, it would be useful to show the 21\,cm-line auto-power spectrum and the 21cm-LAE correlation coefficient of the test datasets. Here it should be noted that the statistical property of training and test data is the same.

\begin{figure*}
\resizebox{160mm}{!}{
\centering
\includegraphics[width=0.5\textwidth,angle=0]{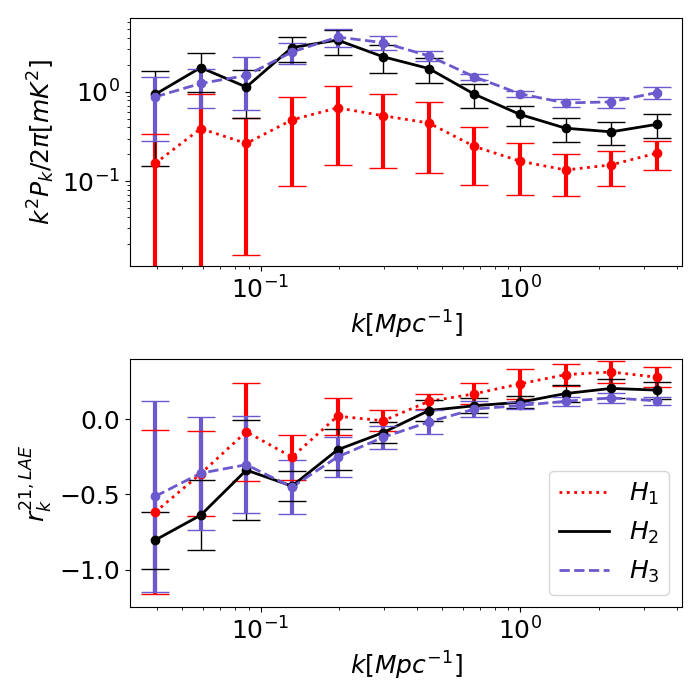}
\includegraphics[width=0.5\textwidth,angle=0]{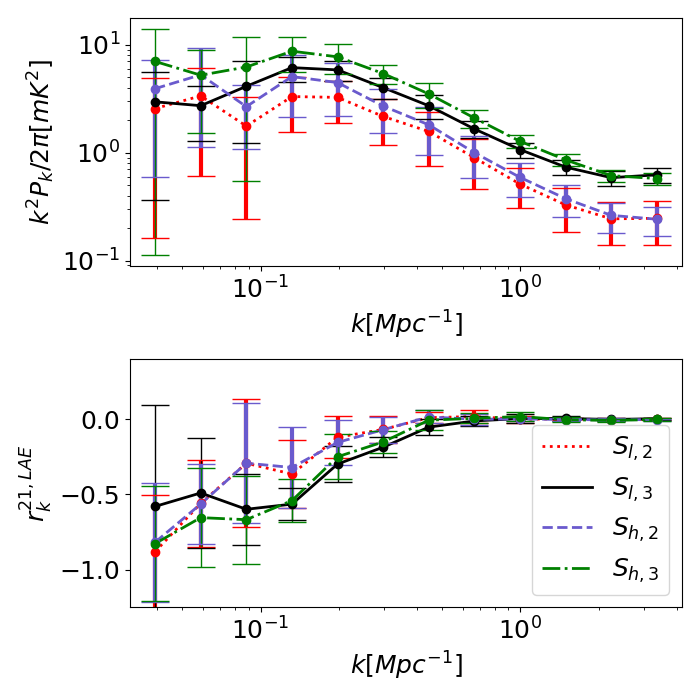}
}
\caption{Comparison of the properties of the test datasets. Top row shows the 21\,cm-line auto-power spectrum, and the bottom row shows the correlation coefficient between the 21\,cm-line signal and LAE distribution. In the left column, dotted line, solid line and dashed line represent $H_{1}$, $H_{2}$ and $H_{3}$ models, respectively. In the right column, dotted line, solid line, dashed line and dot-dashed line are $S_{l,2}$, $S_{l,3}$, $S_{h,2}$ and $S_{h,3}$ models, respectively. The error bars represent the sample variance.}
\label{fig:compmodel} 
\end{figure*}

In upper panels of Fig.~\ref{fig:compmodel}, 21\,cm-line auto-power spectra, $\Delta_{21} \propto k^2 P_{21}$, of 7 training datasets are shown with the standard deviation. As can be seen, they have a peak at $k=0.2~{\rm Mpc^{-1}}$ in all models. Although the power spectra of many models are consistent with each other within sample variance, they show a weak dependence on the model parameters. Since the amplitude of power spectrum is proportional to $x^2_{\rm HI}$, those of high-$x_{\rm HI}$ models ($H_3$, $S_{l,3}$ and $S_{h,3}$) are systematically larger than those of low-$x_{\rm HI}$ models. Besides, although $S_{l}$ models are consistent with $S_{h}$ models at all scales within sample variance, $S_{l}$ models have slightly lower power than $S_{h}$ models at large scale. This indicates that the lower $\alpha_{\rm esc}$ value reduces the ionization contribution from heavier galaxies and mitigates fluctuations due to large ionized bubbles.

The lower panels of Fig.~\ref{fig:compmodel} show the 21cm-LAE cross correlation coefficient, $r_k^{\rm T,O}$. As we indicated before, 21\,cm-line signal and the LAE distribution have a negative correlation at large scales because the inside-out reionization is driven by galaxies in all models. Furthermore, at small scales, the 21\,cm-line correlates positively with LAEs in the $H$ model. This positive correlation is caused by partially neutral regions within the ionized bubbles as shown in \cite{2018MNRAS.479.2754K}. On the other hand, in the $S$ models, the neutral fraction at ionized bubbles is completely zero. Thus, the 21\,cm-line does not correlate with LAEs at $k > 1~\rm Mpc^{-1}$. Although the sample variance is rather large, we can see a tendency that the cross correlation is weaker for lower-$x_{\rm HI}$ models.

\subsection{Image Reconstruction}\label{sec:5.2}

\begin{figure*}
\resizebox{160mm}{!}{
\centering
\includegraphics[width=0.5\textwidth,angle=0]{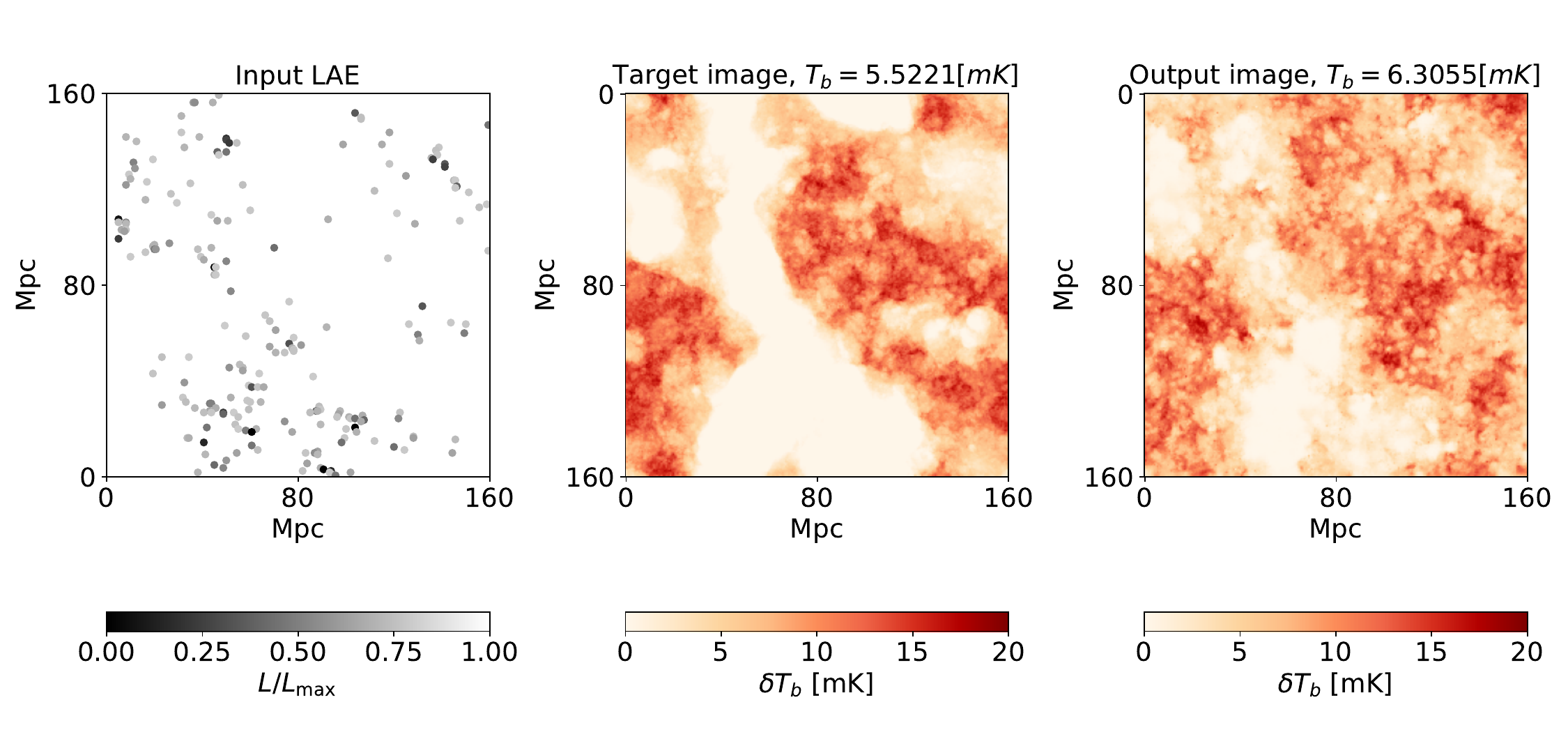}
}
\caption{Example of the 21\,cm-line image reconstruction in $S_{h,3}$ model. The panels show the input LAE distribution, the target 21\,cm-line image, and the output (predicted) image from left to right. In the target image, neutral hydrogen is located far from LAEs and this feature is well learned in the output image. The output image resembles the target image at large scales, the small-scale structure is not well reconstructed.}
\label{fig:fig1} 
\end{figure*}

Fig.~\ref{fig:fig1} shows examples of the input LAE distribution, the target 21\,cm-line image and the predicted 21\,cm-line image for the $S_{h,3}$ model. Here, the predicted image is generated by a network trained with data of $S_{h,3}$ model, that is, the correct model. The neutral regions tend to lie far from LAEs in the predicted image, as they should be, and the large-scale feature is very similar between the target and predicted images. However, the small-scale structure is not reproduced well. This will be due to the weak correlation between LAE distribution and the small-scale fluctuations in 21\,cm-line, as we saw in bottom panels of Fig.~\ref{fig:compmodel}.

In Table~\ref{table:re1}, we summarize the average correlation coefficient between target and predicted images, $r$, and the average number of LAEs for all 14 models. The value of $r$ of $H$ models is generally less than 0.3, while it is $0.4 \sim 0.6$ for $S$ models. This would be due to the larger number of training datasets (see Table~\ref{table:model}), the larger number of LAEs and the simpler distribution of the neutral fraction. It should be noted that the $r$ is a measure of overall similarity between the two images. In fact, as we saw in Fig.~\ref{fig:fig1}, the large-scale feature is reproduced better than the small-scale feature.

\begin{figure*}
\centering
\resizebox{150mm}{!}{
\centering
\includegraphics[width=0.5\textwidth,angle=0]{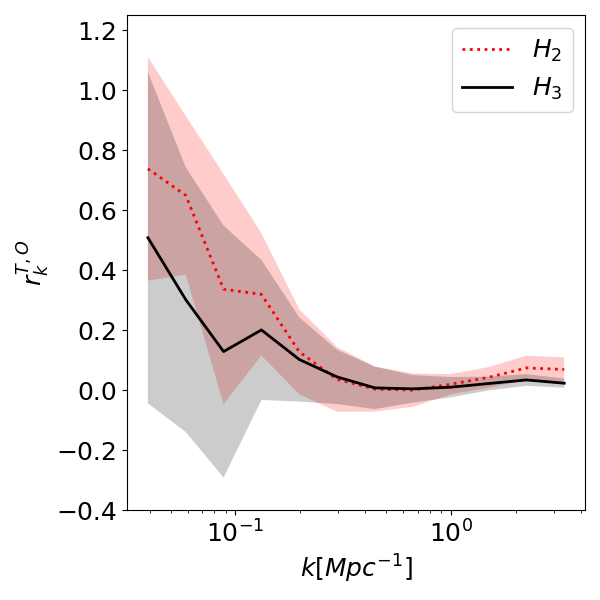}
\includegraphics[width=0.5\textwidth,angle=0]{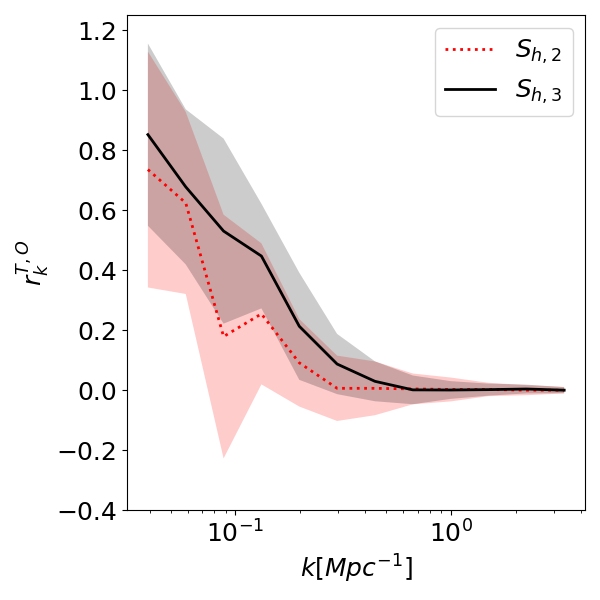}
}
\caption{Correlation coefficient between the target and output 21\,cm-line images. Left and right panels are the result of ($H_2$, $H_3$) and ($S_{2,h}$, $S_{3,h}$) models. The shade shows the sample variance. The cross correlation is relatively strong at large scales ($k < 0.1~{\rm Mpc}$).}
\label{fig:CCe} 
\end{figure*}

To see this quantitatively, in Fig.~\ref{fig:CCe}, we show the scale-dependent correlation coefficient between the two images, $r^{T,O}_k$, defined in Eq.~(\ref{eq:rTO}). The value of $r^{T,O}_k$ is about $0.5$ for $H_3$ model, and $0.8$ for $H_2$ model at the largest scale (left panel), while it is larger than 0.7 for $S_h$ models (right panel) and $S_l$ models (not shown). On the other hand, the correlation is close to zero at small scales ($k > 0.2~{\rm Mpc}^{-1}$) for all models. Nevertheless, looking at the small scales more carefully, $r^{T,O}_k$ for $H_2$ model is non-zero ($\sim 0.1$) at $k \sim 1~\rm Mpc^{-1}$, while it is consistent with zero for $S_h$ models. As we saw in Fig.~\ref{fig:compmodel}, in $H$ models, $r^{\rm 21,LAE}_{k}$ is also positive at small scales and this is caused from the correlation between the LAE distribution and residual HI. \tr{Thus, our network learns the correlation between the 21\,cm-line and LAEs at  large scales, and the network might learn a relationship between the 21\,cm-line and LAEs at the small scales.}

\begin{table*}
  \begin{tabular}{ccccccc}
   \hline \hline
   model &$H_{1}$ & $H_{2}$  &$H_{3}$ & $H_{1}$, UD &$H_{2}$, UD & $H_{3}$, UD  \\
    r & - & 0.28 $\pm$ 0.06 & 0.17 $\pm$ 0.07 & 0.22 $\pm$ 0.08 & 0.39 $\pm$ 0.05 & 0.33 $\pm$ 0.08  \\
$N_{\rm LAE}$ & 209 &167 &72 &586 &483 &231\\
\hline
    model &\,&$S_{l,2}$ & $S_{l,3}$ &\, &$S_{l,2}$, UD & $S_{l,3}$, UD \\
 r && 0.41 $\pm$ 0.13 & 0.42 $\pm$ 0.09 && 0.47 $\pm$ 0.13 & 0.55 $\pm$ 0.08\\
$N_{\rm LAE}$&\, & 221 &163 &\,&558 &436\\
   \hline 
  model &\,&$S_{h,2}$ & $S_{h,3}$  &\,&$S_{h,2}$, UD & $S_{h,3}$, UD \\
 r  && 0.48 $\pm$ 0.15 & 0.49 $\pm$ 0.11 && 0.60 $\pm$ 0.14 & 0.61 $\pm$ 0.12 \\
$N_{\rm LAE}$&\, & 218 &181 &\,&558 &468\\
   \hline
  \end{tabular}
\caption{\label{table:re1}Correlation coefficient, $r$, between the target 21\,cm-line image and the output (predicted) 21\,cm-line image. The average number of input LAEs per image is also listed. For the model named with UD, we assume the ultra deep survey of the HSC, and then the number of LAEs drastically increases and the cross correlation becomes strong. \tb{The image reconstruction fails for the $H_1$ model and the $r$ is incalculable.}}
\end{table*}

Next, we consider the relation between the quality of reproduction and the number of LAEs. As ionized bubbles for $H$ models are smaller than that for $S$ models and the neutral fraction is not zero even in ionized bubbles for $H$ models due to the recombination and ionization equilibrium, neutral hydrogen can survive even around halos. Therefore, although the averaged neutral fraction of $H_i$ and $S_i$ models ($i = 2$ and $3$) are almost the same and they are based on the same halo distribution, there is a large difference in the number of LAEs.

\begin{figure}
\centering
\resizebox{70mm}{!}{
\centering
\includegraphics[width=0.5\textwidth,angle=0]{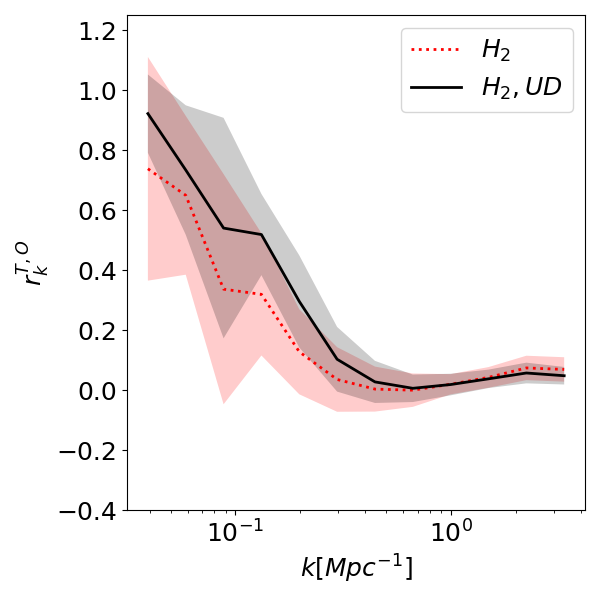}
}
\caption{Same as Fig.~\ref{fig:CCe} but for $H_2$ (dotted) and $H_2,\rm UD$ (solid) models. The ultra deep observation increases the number of LAEs and improves the correlation between the target and output 21\,cm-line images.}
\label{fig:UD} 
\end{figure}

The number of LAEs increases when we assume the ultra deep survey, and the $r$ for the ultra deep survey is larger than that of deep survey as listed in Table~\ref{table:re1}. In Fig.~\ref{fig:UD}, we show the $r^{T,O}_k$ between target and output images for $H_2$ and $H_2, \rm UD$ models. It improves from 0.7 to 0.9 with an assumption of the ultra deep survey. These facts indicate that, as the number of LAEs increases, the relation between the LAE and 21\,cm-line signal is learned more accurately.

\subsection{Statistical Property}\label{sec:5.3}

We now investigate statistical metrics such as the 21cm power spectrum $P_k$ and the 21cm-LAE cross correlation coefficient $r^{21,\rm LAE}_{k}$ of predicted images and compare them with those of target images.

\begin{figure*}
\centering
\resizebox{160mm}{!}{
\centering
\includegraphics[width=0.5\textwidth,angle=0]{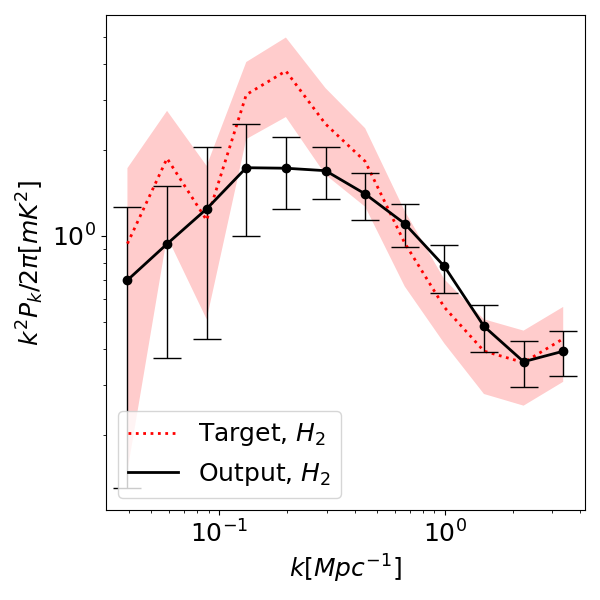}
\includegraphics[width=0.5\textwidth,angle=0]{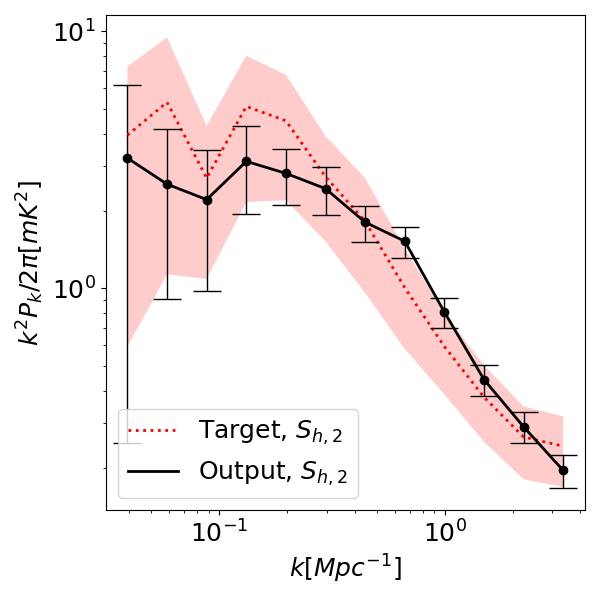}
}
\hspace{10mm}
\resizebox{160mm}{!}{
\centering
\includegraphics[width=0.5\textwidth,angle=0]{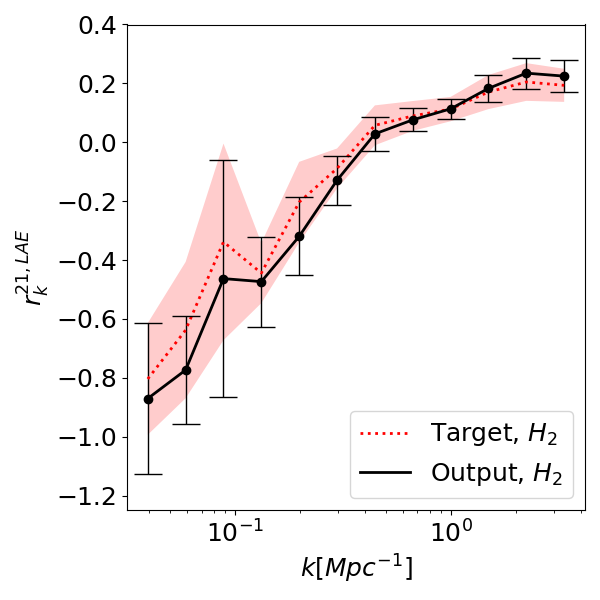}
\includegraphics[width=0.5\textwidth,angle=0]{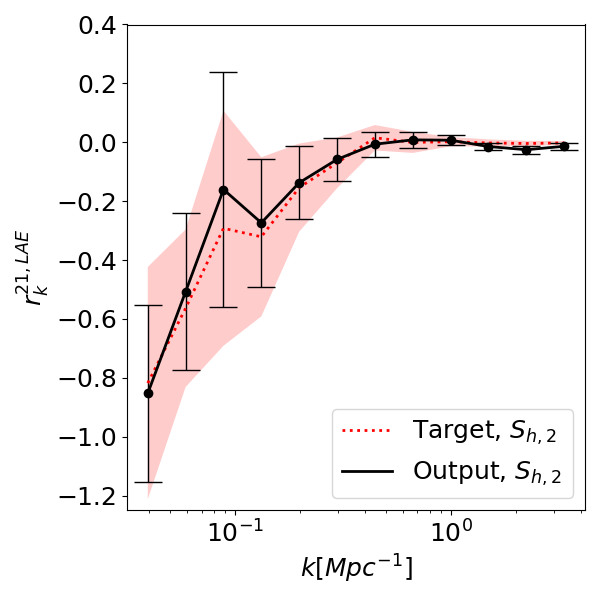}
}
\caption{Top and bottom panels represent the 21\,cm-line auto-power spectra and the cross correlation coefficient between the 21\,cm-line signal and LAEs. Left and right panels show the comparison of target (dotted) and output (solid) images for $H_2$ and $S_{h,2}$ models, respectively. Shade and error bars represent the sample variance evaluated from the target and output samples, respectively. The network succeeds to generate output images with the 21\,cm-line auto-power spectrum and 21cm-LAE cross power spectrum which are consistent with those of the target images. In other models not shown here, the quality of the networks is similar.}
\label{fig:PS1} 
\end{figure*}

In the upper panels of Fig.~\ref{fig:PS1}, we show the 21\,cm-line auto-power spectrum of predicted and target images of $H_2$ and $S_{h,2}$ models. Interestingly, their power spectra are consistent within sample variance even at small scales, although, as we saw in the previous subsection, the correlation between target and predicted images is poor at small scales. This result indicates that the network succeeds to learn the amplitude of 21\,cm-line fluctuations at all scales but fails to learn the phase of fluctuations at small scales. 

The bottom panels of Fig.~\ref{fig:PS1} show the correlation coefficient between the 21\,cm-line map and the LAE distribution, $r^{21,\rm LAE}_k$. As with the case of the 21cm power spectrum, predicted images are consistent with the target images within sample variance. The $P_k$ and $r^{21,\rm LAE}_{k}$ of other models, which are not shown here, have the same accuracy\footnote{\tb{For the $H_3$ and $H_1$, UD models, the accuracy of the reconstruction of $P_k$ is lower than other models.}} as Fig.~\ref{fig:PS1}. Therefore, our network can predict 21\,cm-line images which are statistically consistent with target images.

\subsection{Cross Test}\label{sec:5.4}

So far, we checked the quality of the networks which were trained with the correct models. In fact, we cannot know the correct model a priori. Thus, \tb{as a practical use of the networks, we propose to take a cross correlation between the observed images and predicted 21\,cm-line images of various models, and find the model which has the strongest correlation. Through the analysis, we could identify the best model which may resemble the true physics.} To see how well this works, we here calculate the cross correlation between 21\,cm-line images of $H_2$ model and predicted images obtained by networks trained by correct and different models.

\begin{figure*}
\centering
\resizebox{160mm}{!}{
\centering
\includegraphics[width=0.5\textwidth,angle=0]{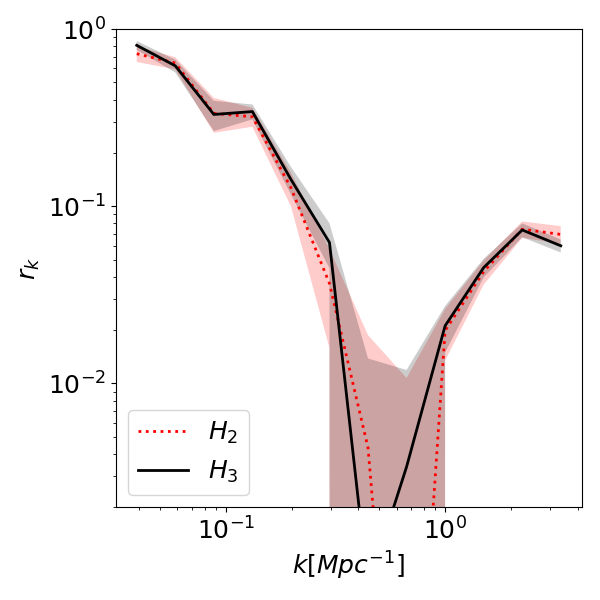}
\includegraphics[width=0.5\textwidth,angle=0]{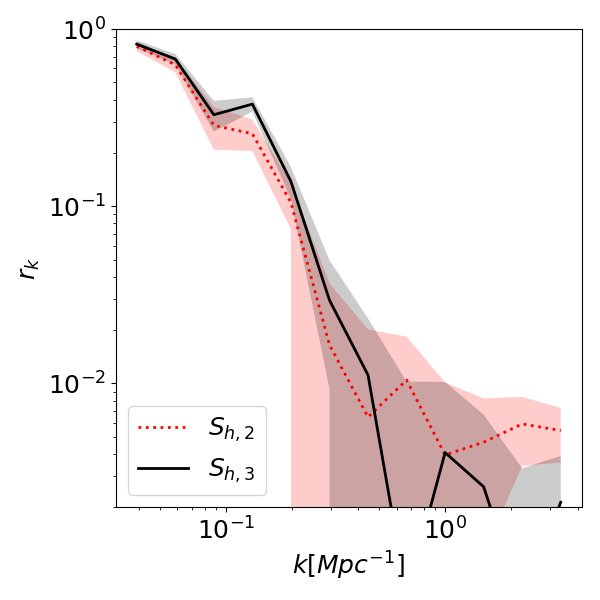}
}
\caption{Correlation coefficient between the predicted images of $H_2$ and other models. The output images are generated from the input LAE distribution of $H_2$ model by the networks trained by 21\,cm-line images of other models. Shades show the sample variance, and the survey area of $27~{\rm deg}^2$ is assumed. It is seen that the cross correlation is measurable at large scales even if the model of the network is not correct.}
\label{fig:cross}
\end{figure*}

In Fig.~\ref{fig:cross}, we show the cross correlation between target images of $H_2$ model and predicted images of other models, $r^{H_2,O}_k$, with a shade representing the sample variance calculated as $\sigma/\sqrt{N}$ where $\sigma$ is the standard deviation, $N$ is the number of samples and we assume $N = 27$ considering the HSC deep survey.

As we see in the left panel, even for the $H_3$ model which has different neutral fraction, the correlation coefficients between the predicted images and the target image of $H_2$ model is relatively high ($\sim 0.5$ for $H_3$) at large scales. Interestingly, the predicted images by $S$ models shown in the right panel correlate even stronger ($\sim 0.8$) with $H_2$-model images at large scales.

Relatively strong correlation for all models at large scales is due to the fact that they are all based on the inside-out reionization scenario. Thus, on the one hand it is not easy to measure neutral fraction and/or distinguish models, but on the other hand the detectability of 21\,cm-line signal by this method does not depend on the detail of the model used to create training data.

Here it should be noted that the correlation, $r^{H_2,O}_k$, for $H$ models is positive at small scales ($k > 1 \rm Mpc^{-1}$) while it vanishes for $S$ models. Thus, observations at small scale may be useful to identify the correct model.

\subsection{Detectability}\label{sec:5.5}

\tb{As one of the practical use of the current method, we propose taking the cross correlation between cGAN-output 21\,cm-line image and observed 21\,cm-line image. In this section, we evaluate the detectability of the cross power spectrum by making mock data of 21\,cm-line observation from the target 21\,cm-line image. We assume the MWA Phase II compact array observation and the HSC deep survey. First, in order to evaluate the intrinsic detectability, we ignore the foreground contamination. Next, we discuss the impact of the foreground residuals. It should be noted that the field-of-view of MWA at $z = 6.6$ is much larger than the HSC deep field. However, in practice, the HSC deep field consists of 4 separate patches of sky with an area of $7 ~\rm deg^2$ per each field. Therefore, we assume MWA observations for four pointings. The total survey area is $27~{\rm deg}^2$, and we assume the observation time of the MWA is 250 hours per HSC deep field. The error is calculated using Eq.~(\ref{eq:error}).  Note that the model used for training is identical to the model of target images in this analysis. 
}

\begin{figure*}
\centering
\resizebox{160mm}{!}{
\centering
\includegraphics[width=0.5\textwidth,angle=0]{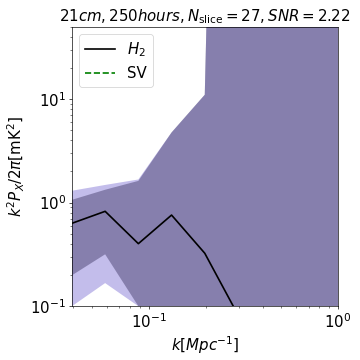}
\includegraphics[width=0.5\textwidth,angle=0]{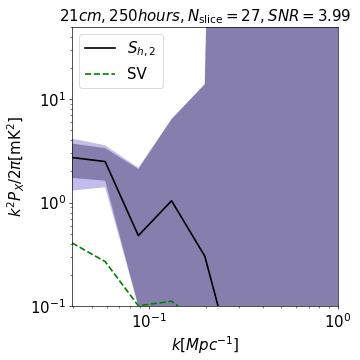}
}
\caption{Detectability of the cross power spectrum between target 21\,cm-line maps and output images in the $H_{2}$ and the $S_{h,2}$ cases. Observation of HSC deep survey with the FoV of 27 $\rm deg^2$ and the MWA Phase II observation of 250 hours per HSC fields are assumed. Solid line is the cross power spectrum, the dashed line is expected sample variance error (SV), the \tb{darker} shade is the error calculated from Eq.~(\ref{eq:error}) including the thermal noise of the MWA and \tb{lighter shade is the error including foreground residual with $f_{\rm rm}=10^{-7}$.} The cross correlation is observable at large scales  ($k<0.1\rm  Mpc^{-1}$) for the $S_{h,2}$ model and the cumulative signal to noise ratio is larger than 3.}
\label{fig:feasibility} 
\end{figure*}

Fig.~\ref{fig:feasibility} shows the cross power spectrum between {target} and predicted 21\,cm-line images for $H_2$ and $S_{h,2}$ cases. The shaded region represents the observational error on the cross power spectrum. The cross power spectrum is detectable at $k < 0.1~{\rm Mpc}^{-1}$ for both models. The cumulative signal-to-noise ratio (SNR), which is calculated as ${\rm SNR^2} = \sum_k (P_X(k) / \sigma_X(k))^2$, \tb{exceeds 3 for the $S_{h,2}$ model.}

\tb{As we consider the cross power spectrum of the 2D images, rather than 3D distribution} the cross power spectrum cannot be detected at small scales since the MWA phase II compact array configuration does not have long baselines sufficient for resolving small-scale fluctuations. The detection at small scales might be possible by the SKA1-Low, although the cross power spectrum is weak at small scales. \tb{However, by considering the three dimensional cross-correlation, the sensitivity at small scales will be improved even for the MWA. Since the number density of the baselines are high at low $k_{\perp}$, the small scale sensitivity can be improved by using the k-modes at high $k_{\parallel}$. To take the information on the line of sight into account, however, the three dimensional image reconstruction is required. The three dimensional LAE distribution can be provided once the precise redshift information of LAEs is obtained with the future PFS spectroscopic observation. Also, since we do not have enough simulation data available for the three dimensional reconstruction, we have focused on the 2D image reconstruction.}

\begin{figure}
\centering
\resizebox{70mm}{!}{
\centering
\includegraphics[width=0.5\textwidth,angle=0]{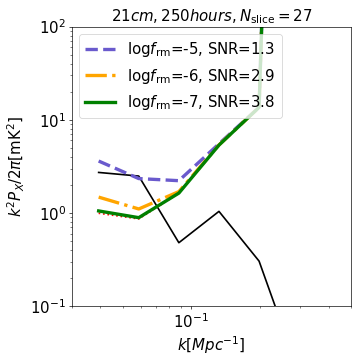}
}
\caption{\tb{Comparing the cross power spectrum and 1 $\sigma$ error. The thin solid line is the cross power spectrum of the $S_{l,2}$ model. The error with foreground residuals are shown as dashed, dot-dashed and solid lines with $f_{\rm rm}=10^{-5}$, $10^{-6}$ and $10^{-7}$, respectively. The total SNR is shown in the label. Dotted line covered by the solid line is the error without foreground contamination. }}
\label{fig:fgrm} 
\end{figure}

\tb{So far, we discussed the intrinsic detectability with an assumption of perfect foreground removal. In fact, we cannot expect a perfect foreground removal and there must be residual foregrounds. Thus, we here discuss the error caused by the foreground residuals. As we stated before, the foreground power spectrum is modeled as Eq.~\ref{eq:fg} and the power spectrum is more than 7 orders of magnitude larger than the 21cm power spectrum. Thus, the statistical error due to the foregrounds dominates the cross power spectrum without foreground removal techniques. We assume the foregrounds are reduced by a factor of $f_{\rm rm}$ at all scales and show the error by varying the $f_{\rm rm}$ in Fig.~\ref{fig:fgrm}. We find that the removal of $f_{\rm rm}=10^{-7}$ is required to detect the cross correlation. \tb{As the error is reduced by increasing the $S_{\rm area}$, the required $f_{\rm rm}=10^{-7}$ becomes lower by widening the LAE survey area.} The level of foreground removal is expected to be achieved by methods such as GMCA \citep{2013MNRAS.429..165C} and GPR \citep{2018MNRAS.478.3640M}.} \tb{We note that since the foreground power spectrum highly depends on the direction of the sky, the power of foreground residual compared to the 21cm line might be a better indicator rather than $f_{\rm rm}$. For our models of foreground and 21cm signal, the residual foreground power spectrum must be less than five times the 21cm-line power spectrum. }

As mentioned above,  the foreground contamination is reduced by integrating independent samples since the foregrounds contribute only to the statistical variance but not to the average. However, the error due to foregrounds cannot be reduced in the same way in the case of the 21\,cm-line auto-power spectrum. Currently, the most serious obstacle of 21\,cm-line auto-power spectrum measurement is the contamination from the combination of foregrounds and various systematics due to instruments and analysis. The cross correlation technique has a potential to mitigate such systematic errors and should be essential to validate the detection of the 21\,cm-line in future observation.

The foreground avoidance method, based on spectral smoothness of synchrotron emission, is not applicable in this work since we only use spatial information. If we extend our method to 3 dimensional map, the foreground avoidance method is applicable and the cGAN construction can be more useful.

\subsection{Estimatimation of Error in 21cm-line Auto-Power Spectrum}\label{sec:5.6}

As discussed before, because we don't know the correct reionization model a priori, we need to seek a model which has the highest correlation with observed 21\,cm-line images. Once the best model is selected, the 21\,cm-line auto-power spectrum can be calculated with the predicted images of the selected model as demonstrated in Fig.~\ref{fig:PS1}. However, since the cross power spectrum could be measured only at large scales, the uncertainty at small scales should be properly propagated to the estimation of the auto-power spectrum.

It is possible to estimate the error in the 21\,cm-line auto-power spectrum as follows, using the measured cross power spectrum between the observed 21\,cm-line and predicted images. First, 21\,cm-line auto-power spectrum can be {evaluated} as,
\begin{eqnarray}
\sqrt{P'_{21}(k)} &=& \frac{P_X(k)}{r_k(k) \sqrt{P^{\rm pre}_{\rm 21}(k)}},
\label{eq:pred}
\end{eqnarray}
\tr{where $P^{\rm pre}_{\rm 21}$ is the auto-power spectrum of predicted image and $P_X$ is cross power spectrum between  observed 21\,cm-line and predicted images.} Therefore, by using a model of $r_k$ denoted as $\tilde{r}_k$, we can estimate $P'_{21}$ from the observed cross power spectrum and auto-power spectrum of predicted image, where \tr{$~'$ indicates estimated values}. We assume that the $\tilde{r}_k$ is empirically obtained from simulations with a statistical error of $\Delta r_k$ which is replaced by the sample variance of $r^{T,O}_k$. Then the error on the estimated power spectrum can be described as 
\begin{eqnarray}
\sigma^2 \approx \left( \frac{\sigma_X^{2}}{P_X^{2}} + \frac{\Delta^2 r_k}{\tilde{r}_k^2} + \frac{\Delta^2 P_{\rm out}}{4P^2_{\rm out}} \right) \frac{P_X}{\tilde{r}_k P^{1/2}_{\rm out}} ,
\label{eq:prederr}
\end{eqnarray}
where $\sigma_X$ is evaluated in Eq.~(\ref{eq:error}). The error on the 21\,cm-line auto-power spectrum of the predicted image $\Delta^2 P_{\rm out}$ is given as the sample variance of $P_{\rm out}$ which is shown in Fig.~\ref{fig:PS1}. We note that the total error is also reduced by $1 / \sqrt{N}$, \tb{where $N=27$ represents the assumed number of survey areas of total $27~{\rm deg}^2$.}

\begin{figure*}
\centering
\resizebox{160mm}{!}{
\centering
\includegraphics[width=0.5\textwidth,angle=0]{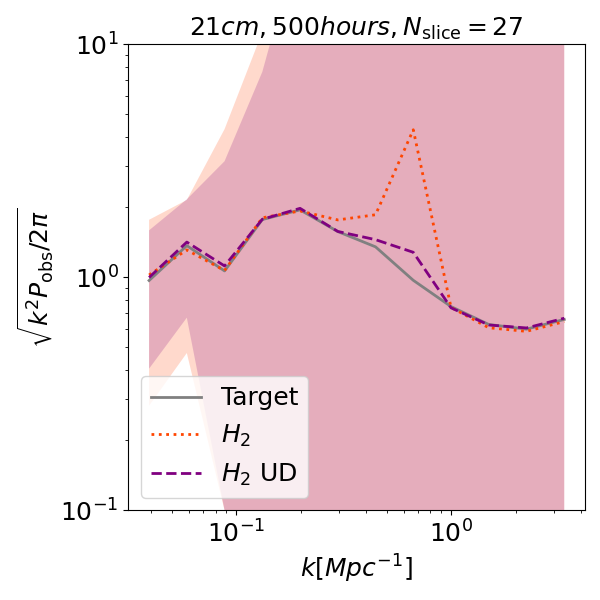}
\includegraphics[width=0.5\textwidth,angle=0]{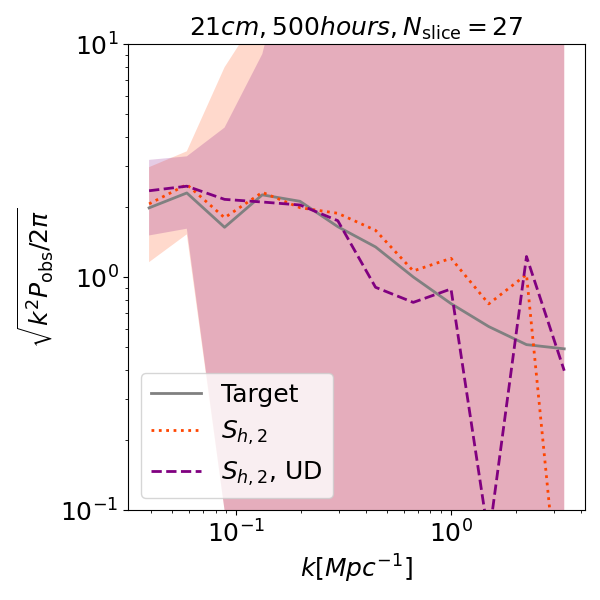}
}
\caption{Prediction of the 21cm power spectrum from the cross power spectrum in the $H_{2}$ and the $S_{h,2}$ model is shown in left and right panels. Same as Fig.~\ref{eq:prederr}, \tb{we assume 500 hours of MWA observation per HSC fields and $N_{\rm slice}=27$.} The solid line is the 21cm power spectrum of target images, the dotted line is the predicted power spectrum and the dashed line is the predicted power spectrum with the assumption of the HSC ultra deep survey. Light colored shade region indicates the error in the case with the HSC deep survey, and the darker shade is that of the ultra deep survey. Here, we use averaged $r^{T,O}_k$ as an empirical correlation coefficient, and therefore the target power spectrum is well consistent with the predicted spectrum. Note that the total survey area of $27~\rm deg^2$ might be an over assumption for the ultra deep survey. }
\label{fig:pred} 
\end{figure*}

Fig.~\ref{fig:pred} shows the 21cm power spectrum estimated by this method. The left and right panels are the results of $H_2$ and $S_{h,2}$ models. The error is large at small scales due to lack of sensitivity of MWA Phase II.

The HSC ultra deep survey can improve the correlation coefficient, and the error on the power spectrum prediction is smaller than that of the deep survey as shown in Fig.~\ref{fig:pred}. However, the improvement is not drastic, and the assumption of field-of-view ($\sim$ 27 $\rm deg^2$) for the ultra deep survey may be too optimistic.

\section{Discussion}\label{Sec:dis} 

\subsection{Improving Network}

There are some strategies to improve the efficiency and the accuracy of the network, which we summarise here: 

\begin{enumerate}
\item \textit{Calibration of training model} - Since we do not know the true model a priori, we need to prepare a large number of networks trained by various models. We are able to reduce the number of networks by model calibration using other observed quantities, and then the learning can be easier than that of this work. For example, we can remove models which are not consistent with observations such as statistical property of LAEs, constraints on the neutral fraction as a function of redshift and Thomson optical depth of cosmic microwave background photons.

\item \textit{Increase the number of independent training datasets} - In this article, we used only one realization of the N-body simulation for the training and test datasets. Furthermore, to avoid training data becoming identical to test data, we used images integrated along x and y directions as training data, and the image integrated along z direction is used as the test data. As a result, the number of training sets is quite smaller than other works applying CNN to 21\,cm-line maps of 1000 training data (e.g. \cite{2018arXiv180502699G,2019arXiv190707787H,2019ApJ...880..110L}). Thus, the accuracy of our estimation would be improved if we could increase the number of original training datasets as many as previous studies. \tb{We note that the GAN method has been used for the study of 21\,cm-line brightness temperature maps in literature. Because their approach is very different from the current one, it is not possible to compare the performance of our method with their work.}

\tb{
\item \textit{Optimizing hyperparameters} - The cGAN method has many hyperparameters which can affect the quality of image reconstruction. The hyperparameters in this work have been chosen based on the result of the random validation described in Sec.~\ref{Sec:NN}. However, some parameters were fixed and were not explored in our validation due to our limited computational resources. Thus, further optimization by increasing the number of varying parameters of the validation could improve image reconstruction.}

\end{enumerate}

\tb{Furthermore, simpler methods such as CNN with U-net architecture might be preferred in the case of relatively small training datasets. Additionally, the accuracy of the construction can be improved using additional input information such as the UV luminosity of the LAEs.} We leave these improvements for future works.

\subsection{Comparison With 21cm-LAE Cross Correlation}

\tb{Here, we compare the detectability of 21\,cm-line signal between the conventional 21cm-LAE cross power spectrum and our new cross correlation introduced in this work. In this work, 1,000 hours of MWA observation is required to achieve the detection of the new cross power spectrum. On the other hand, as discussed in previous works \citep{2014MNRAS.438.2474P,2018MNRAS.479.2754K,2018MNRAS.479.2767Y,2020MNRAS.494..703W}, 1,000 hours of MWA observation is required for detecting the 21cm-LAE cross power spectrum. Thus, the detectability of each cross correlation is comparable. }

{However, it should be noted that, as discussed above, the accuracy of prediction can be boosted by improving the neural network architecture. If a future improved network can enhance the correlation between the target and predicted images at all scales, the detectability of our new cross correlation would become better than the conventional 21cm-LAE cross correlation.}

{Another advantage of our new cross-correlation method is that the 21\,cm-line map and its auto-power spectrum can be estimated without a model uncertainty. In case of the conventional 21cm-LAE cross correlation, we can construct a model of reionization and LAEs which produces a cross-power spectrum consistent with the observation data. Then, the model can predict a 21\,cm-line map and its auto-power spectrum. However, the observed cross-power spectrum cannot narrow down a model and the predicted auto-power spectrum has a model uncertainty.
}

{
On the other hand, in the current case, the network which generates the 21\,cm-line map is trained by simulation datasets based on a specific model of reionization and LAEs. However, once the generated 21\,cm-line map has a correlation with the observed data, the generated map reflects the target 21\,cm-line map independently of the model used to train the network. Thus, the estimation of the 21\,cm-line map and statistics are more direct for the current method.
}

\subsection{Neutral Fraction Map}

We have discussed the image prediction of the 21\,cm-line brightness temperature. However, the cGAN can predict the neutral fraction map, which cannot be measured directly via the 21\,cm-line. The cross correlation between the observed 21\,cm-line and the reconstructed neutral fraction images can be useful to extract the information of ionized regions from the observation of 21\,cm-line, and it will effectively reveal the property of ionizing sources.

\begin{figure*}
\resizebox{160mm}{!}{
\centering
\includegraphics[width=0.5\textwidth,angle=0]{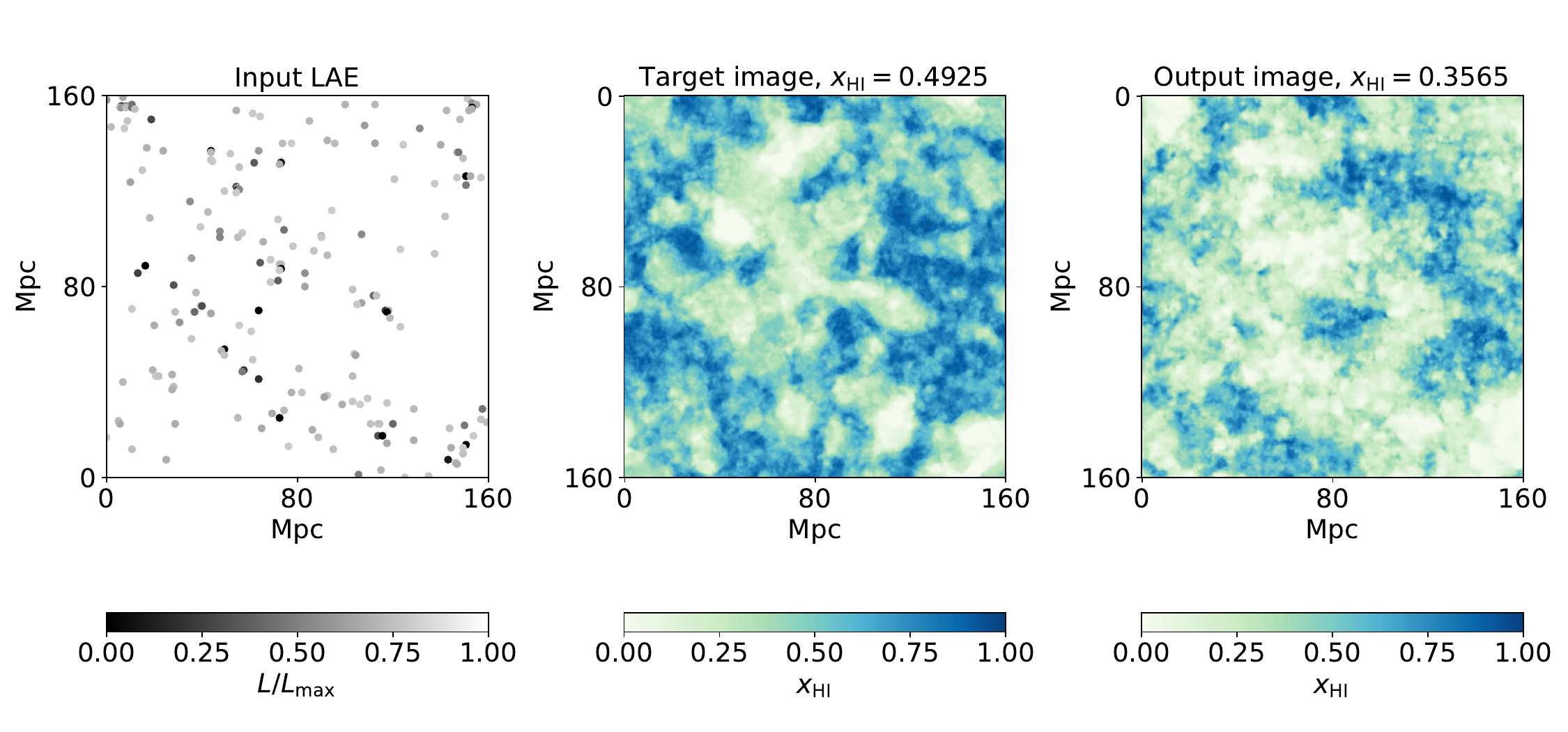}
}
\caption{Example of the reconstruction for the neutral fraction in the $S_{h,3}$ model. The panels show the input LAE distribution, the target image, and the output image from left to right. Same as \ref{fig:fig1}, the large scale structure of the output image is comparable to the target image. 
}
\label{fig:nHI} 
\end{figure*}

\tr{Fig.~\ref{fig:nHI} is an example of the reconstruction for neutral fraction. As in the case of 21\,cm-line, the network can predict the neutral fraction map which resembles the target image.}

Table.~\ref{table:re2} shows the correlation coefficient between the target and predicted neutral hydrogen maps, $r$, its standard deviation and the average number of LAEs, $N_{\rm LAE}$. The result is similar to that of 21\,cm-line which is summarized in table.~\ref{table:re1}.

Here it should be noted that the prediction of averaged $x_{\rm HI}$ was not accurate in this work although the averaged $x_{\rm HI}$ of output images tends to be the correct value of the training dataset. The prediction of the averaged value might require adding the averaged $x_{\rm HI}$ in the optimizing function instead of $L_1$ term.

\begin{table*}
  \begin{tabular}{ccccccc}
   \hline \hline
  model &$H_{1}$ & $H_{2}$  &$H_{3}$ & $H_{1}$, UD &$H_{2}$, UD & $H_{3}$, UD  \\
r & 0.12 $\pm$ 0.10 & 0.26 $\pm$ 0.06 & 0.20 $\pm$ 0.07 & - & 0.39 $\pm$ 0.05 & 0.35 $\pm$ 0.07\\
$N_{\rm LAE}$ & 214 &169 &73 &603 &482 &233\\
  \hline
    model& &$S_{l,2}$ & $S_{l,3}$ & &$S_{l,2}$, UD & $S_{l,3}$, UD  \\
r && 0.43 $\pm$ 0.12 & 0.41 $\pm$ 0.09 && 0.47 $\pm$ 0.11 & 0.56 $\pm$ 0.09  \\
$N_{\rm LAE}$ &\,& 221 &161 &\,&558 &434 \\
   \hline 
  model& &$S_{h,2}$ & $S_{h,3}$ & &$S_{h,2}$, UD & $S_{h,3}$, UD  \\
r && 0.50 $\pm$ 0.15 & 0.50 $\pm$ 0.12 && 0.56 $\pm$ 0.14 & 0.61 $\pm$ 0.11 \\
$N_{\rm LAE}$ &\,& 217 &179 &\,&560 &465 \\
   \hline
  \end{tabular}
   \caption{\label{table:re2}Same as Table.~\ref{table:re1}, but for neutral fraction images. Since we randomly chosen the training sample, the average number of LAEs is not identical to the Table.~\ref{table:re2}. \tb{For the $H_1$, UD, the image reconstruction fails and the $r$ is less than 0.01.}}
\end{table*}

\subsection{Spherical Bubble Model}

In the model of the inside-out reionization, ionized bubbles are generated around galaxies. Although the size and the shape of bubbles depend on the source property and the density of the neutral hydrogen distribution, a spherical bubble (SB) model, in which we assume ionized bubbles can be replaced with spheres around ionizing sources, is a reasonable model at large scales. Thus, the SB model is useful for estimating the map of neutral fraction from distribution of ionizing sources. Note that the SB model might be available for the 21\,cm-line, but the fluctuation of the matter density cannot be taken into account. Thus, here, we use the SB model only for the neutral fraction map.

\begin{figure*}
\centering
\resizebox{160mm}{!}{
\centering
\includegraphics[width=0.5\textwidth,angle=0]{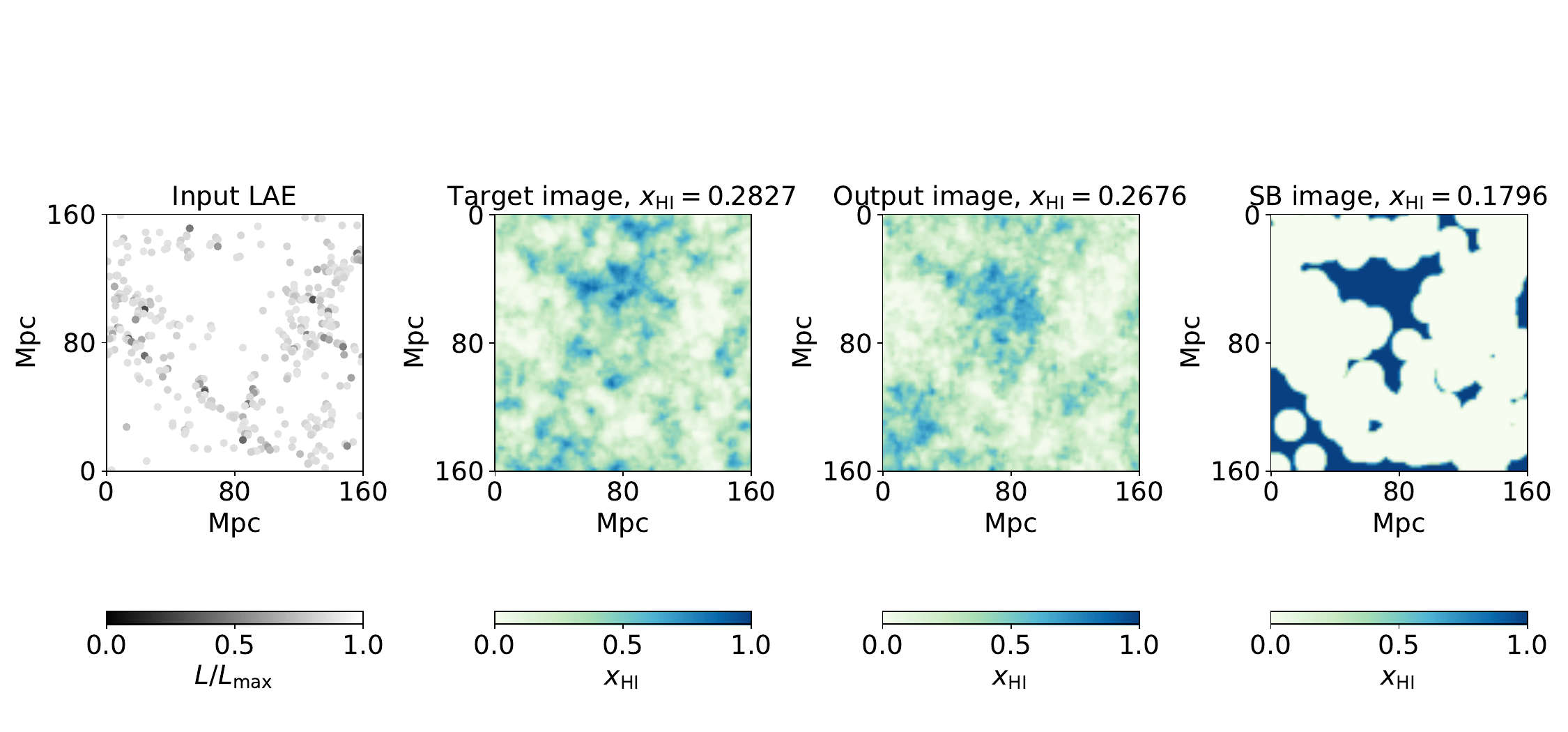}
}
\caption{The panels show examples of the input LAE distribution, the target neutral fraction image, the cGAN output and the SB model images from left to right. This is the $H_{2}, \rm UD$ model. In the SB model, we use $R=16\rm Mpc$. }
\label{fig:SB} 
\end{figure*}

Instead of the cGAN network, we can use the SB model to predict the distribution of the neutral fraction. For the $H_{2},\rm UD$ model, we make ionized regions with the radius $R$ around input LAEs and assume the IGM outside of bubbles is neutral. An example image of LAE distribution, target image, output image and image of the SB model is shown in Fig.~\ref{fig:SB}.

\begin{figure*}
\centering
\resizebox{160mm}{!}{
\centering
\includegraphics[width=0.5\textwidth,angle=0]{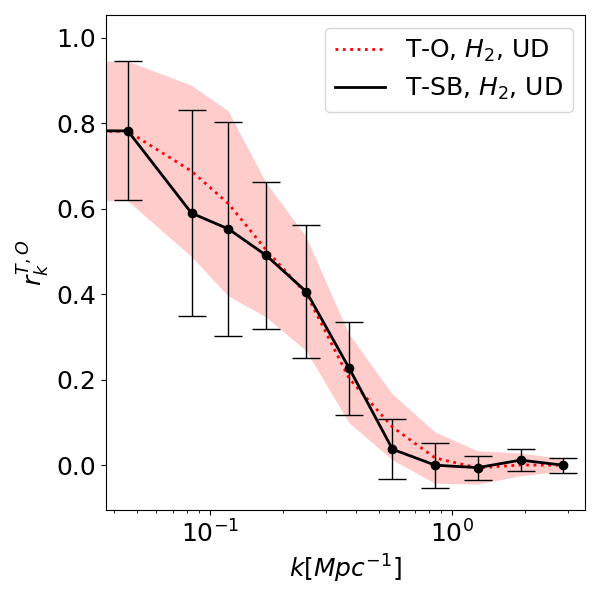}
\includegraphics[width=0.5\textwidth,angle=0]{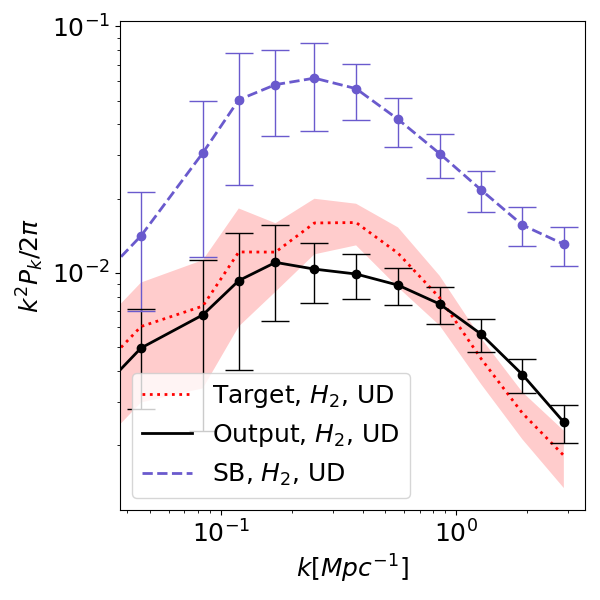}
}
\caption{Left : Dotted line is the correlation coefficient between the target image of the neutral fraction and the cGAN output image, and the solid line is that between the target image and the image in the SB model. We compare the $H_2,\rm UD$ model with the SB model with $R$ of 16 Mpc. The shade and error bar show the sample variance. Right : The dotted, solid and dashed lines are the power spectrum of the neutral fraction for the target image, the cGAN output image and the SB model.}
\label{fig:SB1} 
\end{figure*}

The radius $R$ is a free parameter in the SB model, and we choose $R=16$ Mpc so that the $r_k$ between the target image and cGAN output image is consistent with that of the target image and the SB model as shown in left panel of Fig.~\ref{fig:SB1}. However, if we calibrate the SB model based on the auto power spectrum, $R$ of 22 Mpc is chosen. Since the R is highly ambiguous, the SB model is too simplistic to predict the neutral fraction map, and the cGAN method can be more useful than the SB model.

\section{Summary}\label{Sec:sum} 

In this work, we proposed a new approach to detect EoR 21\,cm-line signal, which measures cross correlation between observed 21\,cm-line map and a 21\,cm-line map predicted from the LAE distribution through machine learning. Specifically, we applied the conditional GAN method to the translation of an LAE distribution map to a 21\,cm-line map. The network is trained using LAE distribution maps, 21\,cm-line brightness temperature maps and neutral fraction maps obtained by the RT numerical simulation ($H$ models) and the semi-numerical simulation ($S$ models). For preparing the LAE distribution, we solved the Lyman-$\alpha$ radiative transfer using the IGM data and models of the line profile.

The network can reproduce the maps of the 21\,cm-line brightness temperature from the LAE distribution. The accuracy depends on the training model and the average neutral fraction. As a metric to assess the quality of networks, we have calculated the cross correlation coefficient between the target and predicted images. Although the overall correlation is rather weak ($0.17 \sim 0.28$ for $H$ models and $0.41 \sim 0.55$ for $S$ models) assuming the HSC deep survey, the correlation at large scales ($k \sim 0.04~{\rm Mpc}^{-1}$) is much better ($0.4 \sim 0.7$ for $H$ models and $0.7 \sim 0.8$ for $S$ models). We also found that the correlation can be stronger by increasing the number of LAEs with the HSC ultra deep survey.

We also used the 21\,cm-line auto-power spectrum and the 21cm-LAE correlation coefficient to evaluate the accuracy of the reconstruction. The result of cGAN output images is consistent with that of target images at all scales within sample variance. This indicates that the network learns the fluctuation in 21\,cm-line signal and the correlation between 21\,cm-line signal and LAE distribution correctly.

\tb{We have evaluated the detectability of the cross correlation between observed 21\,cm-line maps and predicted images. We found that the cross power spectrum can be measured at large scales ($k < 0.1~{\rm Mpc}^{-1}$) with 1000 hours of MWA Phase II observation and the HSC deep survey in assumption of 7 orders of magnitude foreground removal in units of $\rm mK^2$.}

\tr{While we use the cGAN methods to predict the 21\,cm-line from input LAE distributions, the cGAN can be used for other contexts to explore the reionization and the evolution of galaxies at high-$z$. Since the GAN is usually used to generate samples by learning a distribution such as the 21\,cm-line \citep{2020MNRAS.493.5913L}, the LAE distribution also can be created by learning a various models of LAE distribution. The generated samples of the LAEs can be used to compare observation for constraining the astrophysical parameters. }

As a final remark, the cross correlation is a crucial observable to distinguish the 21\,cm-line signal from residuals of foregrounds. Furthermore, if the cross correlation were detected, we have to compare the signal with a number of numerical simulations in order to interpret the result. Our method can confirm the detection and allow us to interpret the reionization scenario simultaneously. Nevertheless, there is room for improvement in our method in many ways: optimizing the network hyper-parameters, increasing the number of training datasets, and adding more information. By boosting the detectability, our method can be an essential tool for future data analysis of the EoR 21\,cm-line observation.

\section*{Acknowledgements}
\tb{We would like to thank anonymous referee for his/her helpful comments.} We also thank Hidenobu Yajima for providing us with the Lyman-$\alpha$ transmission code and Tomoaki Ishiyama for conducting the N-body simulation used in this work. We also thank to Taisuke Nakashima and Kenji Kubota for helpful discussion for the work. SY is supported by JSPS Overseas Research Fellowships. HS was supported by the NSFC (Grant No.11850410429), the China Postdoctoral Science Foundation, the Tsinghua International Postdoctoral Fellowship Support Program, and the International Postdoctoral Fellowship from the Ministry of Education and the State Administration of Foreign Experts Affairs of China. 
This work is partially supported by Grand-in-Aid from the Ministry of Education, Culture, Sports, and Science and Technology (MEXT) of Japan Nos. 21J00416(SY), 	18K03699(KH), 16H05999(KT), 20H00180(KT), 21H01130(KT) and 21H04467(KT). KT is supported by Bilateral Joint Research Projects of JSPS, and the ISM Cooperative Research Program 2020-ISMCRP-2017.

\section*{DATA AVAILABILITY}
The plotting data in this article will be shared on reasonable request to the corresponding author. 




 \bibliographystyle{mnras}
 \bibliography{refbib_}



\appendix
\section{Cross validation}\label{APP1}

\tb{To tune the hyperparameters of the architecture, we performed the validation by varying the parameters as described in Sec.~\ref{sec:3.2}. In table.~\ref{tab:params}, all parameters are listed with each prior distribution and selected value.}

\begin{table}
  \begin{tabular}{ccc}
   \hline \hline
  Parameter & Prior& best value  \\
   $N_{\rm d}$ & [5,6,7,8]& 7  \\
   $\lambda$ & $U$(1,1000) & 519.9 \\
   $f_{\rm g}$ & [16,32,64]& 64   \\
   $f_{\rm d}$ & [16,32,64]& 16   \\
   $b_{\rm s}$ & $U$(8,64)& 11   \\
   $\log Lr$ & $U$(${-8.5}$,${-2.5}$)& -2.55   \\
   $\beta$ & $U$(0.001,0.75)& 0.22   \\
   \hline
  \end{tabular}
 \caption{\label{tab:params}\tb{The cGAN parameters used in this work such as the number of convolution layers $N_{\rm D}$, weight on L1 term $\lambda$, initial number of filters for generator $f_{\rm g}$, initial number of filters for discriminator $f_{\rm d}$, batch size $b_{\rm s}$, initial learning rate for Adam $Lr$ and first momentum term of Adam $\beta$. For the validation, parameters are randomly selected. Here $U(a,b)$ represents a uniform distribution from $a$ to $b$. For $N_{\rm d}$, $f_{\rm g}$, $f_{\rm d}$, we randomly choose the value from the given discrete array.}}
\end{table}


\bsp	
\label{lastpage}
\end{document}